\documentclass[epj,nopacs]{svjour}
\usepackage{latexsym}
\usepackage{graphicx,epstopdf}
\usepackage{epsfig}   
\usepackage[utf8]{inputenc}
\usepackage{amsmath} 
\usepackage{amssymb}
\usepackage{bm}       
\usepackage{color}
\usepackage{slashed}  
\usepackage{hyperref}
\usepackage{physics}  
\usepackage{empheq}

\usepackage[usenames,dvipsnames,svgnames,table]{xcolor}
\hypersetup{colorlinks,linkcolor=MidnightBlue,citecolor=MidnightBlue,urlcolor=MidnightBlue}
\hypersetup{breaklinks=true}
\definecolor{babypink}{rgb}{0.96, 0.76, 0.76}
\definecolor{darkpastelgreen}{rgb}{0.01, 0.75, 0.24}

\begin{document}

\title{Electromagnetic fields and directed flow in large and small colliding systems at ultrarelativistic energies}

\author{Lucia Oliva\thanks{\emph{Present address:} oliva@fias-uni-frankfurt.de} 
}                     
%
%

\institute{Institut f\"{u}r Theoretische Physik, Johann Wolfgang Goethe-Universit\"{a}t, Max-von-Laue-Str. 1, 60438 Frankfurt am Main, Germany}
\date{Received: date / Revised version: date}
%
\abstract{
The hot and dense QCD matter produced in nuclear collisions at ultrarelativistic energy is characterized by very intense electromagnetic fields which attain their maximal strength in the early pre-equilibrium stage and interplay with the strong vorticity induced in the plasma by the large angular momentum of the colliding system. A promising observable keeping trace of these phenomena is the directed flow of light hadrons and heavy mesons produced in symmetric and asymmetric heavy-ion collisions as well as in proton-induced reactions. In particular, the splitting of the directed flow between particles with the same mass but opposite electric charge as a function of rapidity and transverse momentum gives access to the electromagnetic response of medium in all collision stages and in the different colliding systems. The highest influence of electromagnetic fields is envisaged in the pre-equilibrium stage of the collision and therefore a significant imprint is left on the early-produced heavy quarks.
\\
The aim of this review is to discuss the current developments towards the understanding of the generation and relaxation time of the electromagnetic fields embedded in both large and small systems and their impact on the charge-odd directed flow of light and heavy particles, highlighting the experimental results and the different theoretical approaches.
Since it is possible to perform realistic simulations of high-energy collisions that incorporate also the generated electromagnetic fields and vorticity, the study of the directed flow can provide unique insight into the early nonequilibrium phase and the ensuing QGP formation and transport properties.
\PACS{
      {PACS-key}{discribing text of that key}   \and
      {PACS-key}{discribing text of that key}
     } 
} 
\maketitle

\section{Introduction}
\label{intro}

Nuclear collisions at ultrarelativistic energy represent the only laboratory on Earth for investigating the deconfined phase of strongly-interacting matter: the Quark--Gluon Plasma (QGP). Its formation and evolution holds some of the most extreme properties ever observed in nature: 
very high temperature $T$ up to several times the pseudocritical value of the transition between hadronic and partonic matter $T_c\sim155$ MeV $\sim 10^{12}$ K \cite{Bazavov:2014pvz,Ding:2015ona}, i.e. five order of magnitude higher than the temperature at the centre of the Sun; 
very low value of the viscosity over entropy density ratio $\eta/s$ close in the vicinity of $T_c$ to $\eta/s=1/4\pi$ \cite{Romatschke:2007mq,Heinz:2013th}, that is more than twenty times lower than that of the water; 
huge magnetic field up to about $eB\sim50\;m_{\pi}^2\sim10^{19}$ G, i.e. some order of magnitude larger than that expected on the surface of magnetars \cite{Kharzeev:2007jp}; 
intense vorticity up to $\omega\sim 0.1\;c/$fm $\sim10^{22}$ s$^{-1}$ \cite{STAR:2017ckg}, that is 14 orders of magnitude higher than that of any other fluid ever observed.

In the last decades the surprising behaviour of the QGP related to its transport properties has been intensively studied in Heavy-Ion Collisions (HICs).
A new exciting era has begun in recent years after the discovery that small-sized and short-lived droplets of QGP are formed also in small colliding systems; indeed, high-multiplicity events in proton--proton and proton--nucleus collisions at relativistic energy present similar collective features as those found in collisions between two heavy ions.

Both large and small colliding systems are characterized by the presence of extremely intense electromagnetic fields (EMF).
It was realized more than forty years ago that huge magnetic fields could arise in HICs at high energy \cite{Rafelski:1975rf,Voskresensky:1980nk}.
The EMF generated since the the early stage of the collision are mainly due to the spectator protons and in symmetric nucleus-nucleus collisions, such as Au+Au and Pb+Pb, the dominant component is the magnetic field orthogonal to the reaction plane $B_y$.
The first realistic estimates has been carried out in Refs.~\cite{Kharzeev:2007jp,Skokov:2009qp,Voronyuk:2011jd}.
In asymmetric systems, due to the different number of protons in the two colliding nuclei, a remarkable electric field along the impact parameter axis $E_x$ directed from the heavier nucleus towards the lighter nucleus is generated in the central collision region. This has been demonstrated and studied for Cu+Au reactions in Refs.~\cite{Voronyuk:2014rna,Deng:2014uja}.
This asymmetry in the EMF profiles is taken to its extreme in the case of proton-nucleus collisions, where the EMF are almost completely determined by the heavy nucleus and the produced electric field $E_x$ is comparable to $B_y$, as shown and investigated in Ref.~\cite{Oliva:2019kin} for p+Au collisions.
However, event-by-event fluctuations of the proton positions in the colliding ions leads to fluctuations of the EMF and also the other components could reach values of the same order of magnitude of $B_y$ and $E_x$ \cite{Bzdak:2011yy,Deng:2012pc,Toneev:2012zx}.
\\
The initial values of the magnetic field attained in peripheral HICs at top RHIC energy are $eB_y\approx 5 m_{\pi}^2$ and about one order of magnitude higher at LHC energies. However, the decay rate of this huge fields with time is still under debate, being it dependent on the formation timescale and the electromagnetic response of the medium created after the collision.
Indeed, semi-analytic calculations of the magnetic and electric field evolution in a plasma with constant electric conductivity \cite{Tuchin:2013ie,Tuchin:2013apa,Gursoy:2014aka} indicate a strong slowdown of the field decay with respect to the evolution in the vacuum. This description is not reliable in the very early stage, before any medium is produced, but suggests that, after dropping by some order of magnitude, as soon as the conducting matter is created the EMF freezes out in it and lasts as long as the QGP lifetime.
This behaviour is qualitatively in agreement with microscopic simulations where the EMF are dynamically generated from spectator and participant protons as well as newly produced charged particles \cite{Voronyuk:2011jd,Toneev:2016bri}.

The electric and magnetic fields act as accelerators for the charges present in the expanding fireball and could modify the dependence on rapidity $y$ and transverse momentum $p_T$ of the final particle distribution with respect to the reaction plane \cite{Poskanzer:1998yz}:
\begin{equation}\label{eq:part_distr}
\begin{split}
&E \dfrac{d^3N}{dp^3}=\dfrac{d^3N}{p_T dp_T dy d\phi}=\dfrac{d^2N}{p_T dp_T dy}\times\\
\dfrac{1}{2\pi}&
\left[1+2\sum_{n=1}^\infty v_n(p_t,y)\cos\left(n\left(\phi-\Psi_n\right)\right)\right],
\end{split}
\end{equation}
where $\phi$ is the azimuthal angle of the particle momentum $p$ and $\Psi_n$ is the $n$th-order event plane of the collision.
\\
Among the Fourier coefficients of the azimuthal particle distribution in Eq.~\eqref{eq:part_distr} the directed flow $v_1$ is the most sensitive observable to the EMF.
Indeed the $v_1$, given by
\begin{equation}
v_1=\langle\cos\left(\phi-\Psi_1\right)\rangle
\end{equation}
where the brakets $\langle\cdots\rangle$ indicate the average over the particles in the collision, is related to a collective sidewards deflection of particles on the event plane with respect to the beam axis ($v_1\sim p_x/p_T$) and the geometry of EMF and fluid velocities requires that the Lorentz force produces a net push of charges along the $x$ axis with opposite directions for positively and negatively charges. Hence, the splitting of the directed flow between particles with the same mass and different electric charge gives direct access to the collective electromagnetic response of the medium \cite{Gursoy:2014aka,Voronyuk:2014rna,Das:2016cwd,Toneev:2016bri,Chatterjee:2018lsx,Gursoy:2018yai,Coci:2019nyr,Inghirami:2019mkc,Oliva:2019kin,Oliva:2020new}.
Furthermore, the measurements of the $v_1$ produced in different colliding systems as well as for different energies and centralities \cite{Adamczyk:2011aa,Abelev:2013cva,Adamczyk:2014ipa,Adamczyk:2016eux,Adamczyk:2017nxg,Acharya:2019ijj} give the opportunity to theoretical models to test their description of the initial state of high-energy collision, especially in view of the combined influence of EMF and vortical dynamics on the directed flow with a different sensitivity for bulk light particles and early-produced heavy quarks \cite{Bratkovskaya:2004ec,Gursoy:2014aka,Voronyuk:2014rna,Das:2016cwd,Toneev:2016bri,Chatterjee:2017ahy,Nasim:2018hyw,Chatterjee:2018lsx,Gursoy:2018yai,Coci:2019nyr,Inghirami:2019mkc,Oliva:2019kin,Oliva:2020new}.
Besides the EMF and the vorticity, the early stage is affected by a nontrivial pre-equilibrium dynamics \cite{Oliva:2019plz} which have an impact especially on the heavy-quark propagation \cite{Liu:2019lac}.
Being possible to embed both QGP and heavy-flavour dynamics in realistic descriptions of high-energy collisions that include the generated EMF and vorticity and take into account the pre-equilibrium effects, further theoretical developments and improved experimental precision could help to solve the current tension between theory and experiment in particular at the LHC energies.

The aim of the present article is to discuss the directed flow observable as a probe of the EMF produced in large and small colliding systems at high energy.
The various theoretical modelling of the generated EMF are discussed, analysing the spacetime profiles of the fields in the different systems, from symmetric Au+Au and Pb+Pb collisions through the asymmetric Cu+Au to p+Au reactions.
The experimental measurements are reviewed along with the theoretical predictions and investigations for what concern the directed flow of light and heavy hadrons, that constitutes a gold probe of the collective response of the medium to the EMF as well as of the EMF themselves.
Indeed, the charge-dependent $v_1$ could shed light on the timescales of the EMF decay and of the formation of electric charges (i.e. $q\overline{q}$ pairs) in the early pre-equilibrium phase of relativistic collision.
Furthermore, a detailed understanding of the (electro-)magnetic fields in HICs is of great importance for the study of other effects driven by them, such as the Chiral Magnetic Effect (CME) and related transport phenomena \cite{Huang:2015oca,Kharzeev:2015znc}, the splitting in the polarization of hyperons and anti-hyperons \cite{Becattini:2016gvu,Han:2017hdi}, the early-time emission of photons and dileptons \cite{Basar:2014swa,Tuchin:2014hza}, the Schwinger particle production \cite{Tuchin:2013ie,Sheng:2018jwf}.
Moreover, the presence of a magnetic field affects also the QCD phase diagram, modifying the behaviour of the pseudocritical temperature and the position of the critical endpoint \cite{Ruggieri:2014bqa,DElia:2018xwo}.

\section{Electromagnetic fields in high-energy nuclear collisions}
\label{sec:emf}

In order to understand the theoretical and experimental results reviewed in the next sections, it is instructive to discuss the spatial profiles and the temporal evolution of the EMF generated in noncentral HICs as well as in proton-induced reactions, examining the different approaches used for their calculation.

\subsection{Different approaches to compute the EMF}
\label{sec:emf_approaches}

The Maxwell equations for the electric field ${\bm E}$ and the magnetic field ${\bm B}$ in presence of a charge density $\rho$ and a current density $J$ reads:
\begin{equation}\label{eq:Maxwell}
\begin{split}
\nabla\cdot{\bm E} = \rho, & \qquad \nabla\cdot{\bm B} = 0, \\
\nabla\times{\bm E} = -\partial_{t}{\bm B}, & \qquad \nabla\times{\bm B} = \partial_{t}{\bm E}+{\bm J},
\end{split}
\end{equation}
where we imposed $\epsilon=\mu=1$, i.e., the polarization and magnetization response of matter is disregarded.
In the static case one obtains from Eqs.~\eqref{eq:Maxwell} the Coulomb and Biot-Savart laws, but the general solution determines the time-varying fields ${\bm E}({\bm r},t)$ and ${\bm B}({\bm r},t)$ for nonstatic sources $\rho({\bm r},t)$ and ${\bm J}({\bm r},t)$.
\\
Eqs.~\eqref{eq:Maxwell} can be solved by expressing the electric and magnetic fields in terms of the scalar potential $\Phi$ and the vector potential ${\bm A}$
\begin{equation}\label{eq:potentials}
{\bm E}=-\nabla\Phi-\partial_t{\bm A}, \qquad {\bm B}=\nabla\times{\bm A},
\end{equation}
then obtaining from Eqs.~\eqref{eq:Maxwell} inhomogeneous wave equations for the electromagnetic potentials with source terms. These equations can be solved for specified sources, such as continuous distributions of charge and current densities or a single arbitrarily moving point-like charge.
In the latter case one arrives to the famous Liénard-Wiechert potentials, that can be inserted in Eqs.~\eqref{eq:potentials} in order to find the retarded electric and magnetic fields produced by a point-like source with charge $e$ at position ${\bm r}(t)$ with velocity ${\bm v}(t)$:
\begin{equation}\label{eq:LWfields}
\begin{split}
{\bm E}({\bm r},t)=&\dfrac{e}{4\pi}
\left\lbrace
\dfrac{\hat{{\bm R}}-{\bm\beta}}{\kappa^3\gamma^2R^2}
+\dfrac{\hat{{\bm R}}\times\left[\left(\hat{{\bm R}}-{\bm\beta}\right)\times\dot{{\bm \beta}}\right]}{\kappa^3cR}
\right\rbrace_{\mathrm{ret}} \\
&{\bm B}({\bm r},t)=\left\lbrace \hat{{\bm R}}\times{\bm E}({\bm r},t) \right\rbrace_{\mathrm{ret}}
\end{split}
\end{equation}
where ${\bm R}={\bm r}-{\bm r}'$ with ${\bm r}'\equiv{\bm r}(t')$ is the relative position, ${\bm\beta}={\bm v}/c$ and $\dot{{\bm\beta}}=\mathrm{d}{\bm\beta}/\mathrm{d}t$ are related respectively to velocity and acceleration of the particle, $\gamma=(1-\beta^2)^{-1/2}$ is the Lorentz factor and $\kappa=1-\hat{{\bm R}}\cdot{\bm\beta}$; all quantities inside the braces labelled with ``ret'' are evaluated at the times $t'$ that solves the retardation equation $t'-t+{\bm R}(t')/c=0$.
We see that for a point-like charge the magnetic field is always perpendicular to the electric field and to the direction $\hat{{\bm R}}$ from the retarded point.

The EMF act on the propagation of a particle with charge $q$ through the Lorentz force\footnote{We note that throughout the paper we indicate as ``Lorentz force'' both terms depending on the electric and the magnetic field; in the literature this nomenclature is also used for referring to only the magnetic part.}:
\begin{equation}\label{eq:lorentz}
F_{em}=\left(\dfrac{\mathrm{d}{\bm p}}{\mathrm{d}t}\right)_{em}=q\left({\bm E}+{\bm\beta}\times{\bm B}\right).
\end{equation}

In principle, thanks to the superposition principle, the retarded fields \eqref{eq:LWfields} and the Lorentz force \eqref{eq:lorentz} permit a consistent computation of the EMF produced in HICs from a microscopic point of view, by considering the time-dependent fields induces by all present charges (spectators, participants, newly produced particles), taking into account the propagation of the charges themselves in the EMF and the back-reaction of particles on the fields. 
\\
The temporal evolution of the fields computed in this way would account naturally for the electric conductivity $\sigma_{el}$ of the system, intended as the proportionality constant between the electric current ${\bm J}$ induced in the system by the Lorentz force and the Lorentz force itself per unit charge:
\begin{equation}\label{eq:sigma}
{\bm J}=\sigma_{el}\left({\bm E}+{\bm\beta}\times{\bm B}\right).
\end{equation}
Since in many physical cases the velocity of the charges is small, the second term can be neglected and Eq.~\eqref{eq:sigma} is reduced to the well-know Ohm law
\begin{equation}\label{eq:Ohm}
{\bm J}=\sigma_{el}{\bm E}.
\end{equation}
However, in relativistic HICs the particle velocity is high enough (at least along the beam direction) to produce a significant contribution of the magnetic term in Eq.~\eqref{eq:sigma}.
Moreover, the dominant directions of the magnetic and electric fields in HICs are such that the two terms in Eq.~\eqref{eq:sigma} are opposite and there is a partial cancellation.
\\
This means that it is important to consider in the theoretical modelling a scenario as close as possible to the physical situation, depending on the observables under study.
For the topic discussed in this paper, i.e. the effect of EMF on the directed flow in HICs, it is not realistic to consider only the magnetic field $B_y$, since it produces by Faraday induction a significant $E_x$ (to be added to the Coulomb contribution in the asymmetric collision case); moreover, especially at LHC energy, it is plausible that the electric conductivity of the QGP is lower than that given by Eq.~\eqref{eq:Ohm} and usually studied in theoretical models \cite{Cassing:2013iz,Puglisi:2014sha,Soloveva:2019xph} and lattice QCD calculations \cite{Brandt:2012jc,Brandt:2015aqk,Amato:2013naa,Aarts:2014nba}, because of the magnetoresistance, i.e. the contribution of the second term in \eqref{eq:sigma}. This would in turn have an influence on the relaxation time of the magnetic field.
\\
We note that the electric conductivity is not the full story but other contributions to the charged currents produced in HICs could come from other phenomena, such as the CME, and the corresponding conductivities should be considered. The effect of a chiral conductivity $\sigma_{\chi}$ on the EMF has been recently studied in Ref.~\cite{Li:2016tel} through analytic calculations and in Ref.~\cite{Astrakhantsev:2019zkr} by means of lattice simulations.

We see from Eq.~\eqref{eq:LWfields} that retarded EMF from moving charges are constituted by two contributions: the first term is independent from the acceleration and corresponds basically to elastic Coulomb fields decaying for large distance as $R^{-2}$ (``velocity fields''); the second term depends linearly on the acceleration and represents radiation fields varying for large distances as $R^{-1}$ (``acceleration fields'') \cite{Landau:1975}.
Since full computations in the time-dependent case are very complicated, the acceleration fields in Eqs.~\eqref{eq:LWfields} are often neglected for practical calculations; the remaining term corresponds to the field produced by a charge in uniform motion.
Then, the total electric and magnetic fields generated in nuclear collisions are a superposition of the fields generated from all moving charges $q_i$:
\begin{equation}\label{eq:LWfields_simpl}
\begin{split}
e{\bm E}({\bm r},t)=\sum_i
\dfrac{\mathrm{sgn}(q_i)\alpha_{em}{\bm R}_i(t)(1-\beta_i^2)}{\left\lbrace\left[{\bm R}_i(t)\cdot{\bm\beta}_i\right]^2+R_i(t)^2\left(1-\beta_i^2\right)\right\rbrace^{3/2}}, \\
e{\bm B}({\bm r},t)=\sum_i
\dfrac{\mathrm{sgn}(q_i)\alpha_{em}{\bm\beta_i}\times{\bm R}_i(t)(1-\beta_i^2)}{\left\lbrace\left[{\bm R}_i(t)\cdot{\bm\beta}_i\right]^2+R_i(t)^2\left(1-\beta_i^2\right)\right\rbrace^{3/2}},
\end{split}
\end{equation}
Even though with some caution due to the omission of the acceleration terms, Eqs.~\eqref{eq:LWfields_simpl} allow a consistent description of the medium+fields evolution if charge and current distributions are computed dynamically so that the back-reaction of particles on fields is taken into account.
\\
This approach is adopted in the Parton-Hadron-String Dynamics (PHSD) framework.
PHSD is a covariant dynamical approach for strongly interacting many-body systems formulated on the basis of off-shell transport equations which determine the temporal evolution of the system both in the partonic and in the hadronic phase \cite{Cassing:2009vt}. It allows to describe the dynamical evolution of large and small colliding systems at relativistic energies in terms of the microscopic degrees-of-freedom, explicitly accounting for the transition between QGP and hadronic phase.
\\
The EMF are included in the PHSD model in a dynamical manner \cite{Voronyuk:2011jd,Toneev:2012zx}: the formation and evolution of the retarded electric and the magnetic fields are determined by Eqs.~\eqref{eq:LWfields_simpl} with the sum running over all charged particles, considering spectators and participants protons as well as newly produced charged hadrons, quarks and antiquarks (in PHSD the distinction between participants and spectators is dynamical, being the latter those nucleons which did not undergo initial hard scatterings).
The quasiparticle propagation in the EMF is determined by the Lorentz force \eqref{eq:lorentz}. The fields are calculated time-by-time (including the contributions from all charges) so that the back-reaction of particle dynamics on the EMF is taken into account.
PHSD has been used for studying the influence of the EMF on the directed flow in asymmetric HICs and proton-induced collisions.

According to Faraday's law a strongly decreasing magnetic field induces an electric field circulating around the direction of the magnetic field. After the early-times evolution, when the system is in the QGP stage, the electric conductivity of the medium is not negligible and the induced electric field generates an electric current that creates a magnetic field which according to the Lenz rule opposes to the decrease of the magnetic field itself, i.e., points towards the $y$ direction of the initial $B$. 
This collective response of the produced matter at later stages is taken into account naturally in the PHSD approach by solving consistently the Maxwell equations and the generalized transport equations with the inclusion of the Lorentz force.
However, since Eqs.~\eqref{eq:LWfields_simpl} are obtained neglecting the acceleration term in Eqs.~\eqref{eq:LWfields}, the medium electromagnetic response that can lead to a slowdown of the decrease of the magnetic field may be underestimated.

\begin{figure}[t!]
\centering
\includegraphics[trim={0 0 0 0},clip,width=\columnwidth]{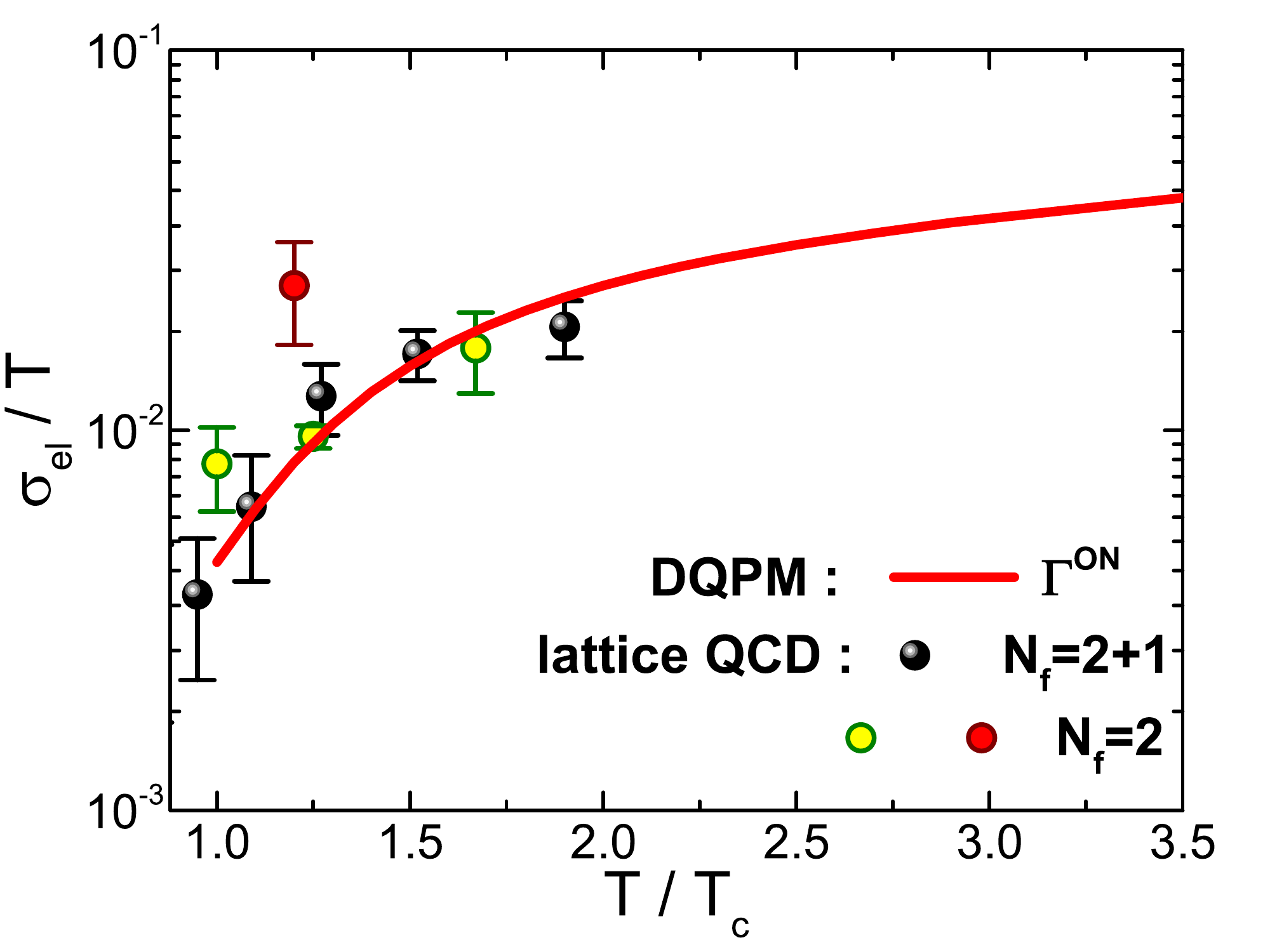}
\caption{(Color online) Ratio of electric conductivity over temperature $\sigma_{el}/T$ as a function of the scaled temperature $T/T_c$. The solid red line is the DQPM results within the relaxation time approximation using the parton interaction rate $\Gamma_i(\mathbf{p},T,\mu)$ for the inverse relaxation time. The symbols are lattice QCD data for $N_f=2$ taken from Refs.~\cite{Brandt:2012jc,Brandt:2015aqk} (red and yellow circles) and for $N_f=2+1$ taken from Refs.~\cite{Amato:2013naa,Aarts:2014nba} (black spheres).
Figure adapted from Ref.~\cite{Soloveva:2019xph}.}
\label{fig:sigmael_T}
\end{figure}

The QGP electric conductivity $\sigma_{el}$ as response of the system to an external electric field has been widely studied in the stationary limit, e.g., in Refs.~\cite{Cassing:2013iz,Puglisi:2014sha,Soloveva:2019xph,Brandt:2012jc,Brandt:2015aqk,Amato:2013naa,Aarts:2014nba}.
In Fig.~\ref{fig:sigmael_T} we show the result of Ref.~\cite{Soloveva:2019xph} for the ratio $\sigma_{el}/T$ as a function of the scaled temperature from the Dynamical QuasiParticle Model (DQPM) included in PHSD approach for defining the QGP properties and interactions on the basis of partonic propagators with sizeable imaginary parts of the incorporated self-energies.
The solid red line is the DQPM result within the relaxation time approximation using the on-shell interaction rate $\Gamma^{on}$. The ratio $\sigma_{el}/T$ rises quadratically with temperature above the critical value $T_c$ which can be related to the increasing number of quarks at higher temperatures.
We note a good agreement of the DQPM result with the available lattice calculations \cite{Brandt:2012jc,Brandt:2015aqk,Amato:2013naa,Aarts:2014nba}.
The electric conductivity can be also computed by solving the relativistic transport equations for the partonic matter in a box with periodic boundary conditions in the presence of an external electric field, as done in Refs.~\cite{Cassing:2013iz,Puglisi:2014sha}.

The electric conductivity is a the main ingredient of another method used for computing the EMF produced in HICs.
It consists in solving analytically the Maxwell equations for a point-like charge $e$ located at the position ${\bm x'}_\perp$ in the transverse plane and travelling along the positive $z$ direction with velocity $\beta$ in a medium with electric conductivity $\sigma_{el}$.
This is done substituting in Eqs.~\eqref{eq:Maxwell} $\rho=\rho_{ext}$ and ${\bm J}={\bm J}_{ext}+{\bm J}_{ind}$, where $\rho_{ext}=e\delta(z-\beta t)\delta({\bm x}_\perp-{\bm x'}_\perp)$ and ${\bm J}_{ext}=\hat{z}\beta e\delta(z-\beta t)\delta({\bm x}_\perp-{\bm x'}_\perp)$ are the external charge and current due to the longitudinally-moving protons and ${\bm J}_{ind}$ is the current induced in the medium and given by the Ohm's law \eqref{eq:Ohm}.
Then, one gets the following wave equations for the electric and magnetic fields:
\begin{equation}\label{eq:wave}
\begin{split}
\left(\nabla^2-\partial^2_t-\sigma_{el}\,\partial_t\right){\bm B} &= -\nabla\times{\bm J}_{ext} ,\\
\left(\nabla^2-\partial^2_t-\sigma_{el}\,\partial_t\right){\bm E} &= -\nabla \rho_{ext}+\partial_t{\bm J}_{ext},
\end{split}
\end{equation}
which can be solved through the method of Green functions in order to find the EMF at an arbitrary spacetime point $(t,z,{\bm x}_\perp)$.
The analytic solution is achievable as the electric conductivity is treated as a constant in all space and time.
The total EMF in the collision are evaluated considering the fields generated by all spectator and participant protons in the two colliding nuclei, while $\sigma_{el}$ accounts for the conducting QGP created after the collision.
\\
This approach has been proposed, with some differences, in Refs.~\cite{Tuchin:2013ie,Tuchin:2013apa,Gursoy:2014aka} and is adopted in further studies for describing the effect of the EMF on final observables; in particular Refs.~\cite{Gursoy:2014aka,Das:2016cwd,Gursoy:2018yai,Chatterjee:2018lsx} have addressed the topic of the EMF influence on particle directed flow in HICs.
However, the assumption of a constant conductivity during all the evolution of the heavy-ion collision may lead to overestimate the slowdown of the magnetic field decay, especially in the early stage when the medium is not yet produced. An improvement would consist in considering the temperature dependence of $\sigma_{el}$ depicted in Fig.~\ref{fig:sigmael_T}.

\subsection{Spacetime profiles of the electromagnetic fields in large and small systems}
\label{sec:emf_profiles}

\begin{figure*}[t!]
\centering
\includegraphics[trim={0 0 0 0},clip,width=0.95\columnwidth]{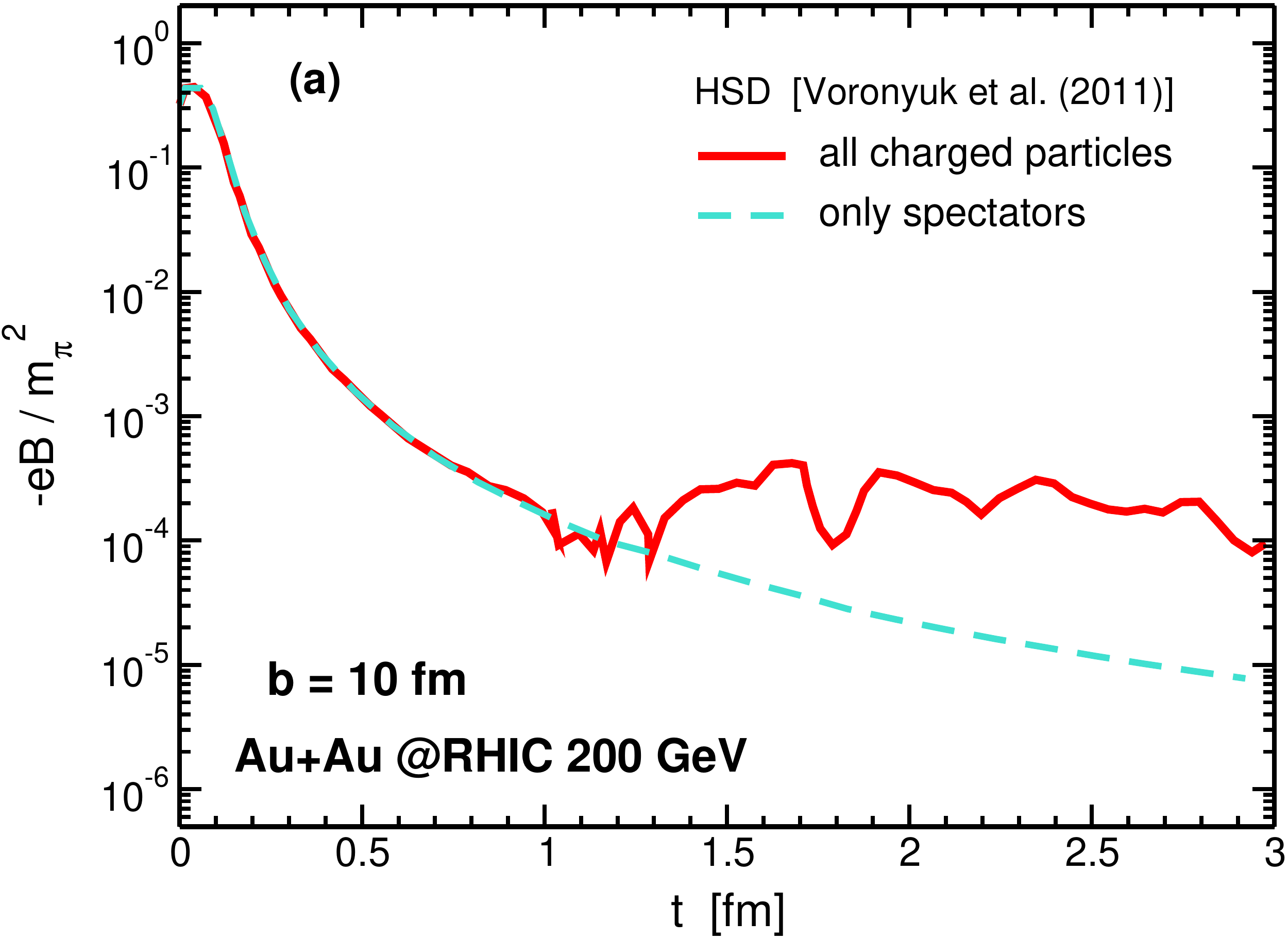}\qquad
\includegraphics[trim={0 0 0 0},clip,width=0.95\columnwidth]{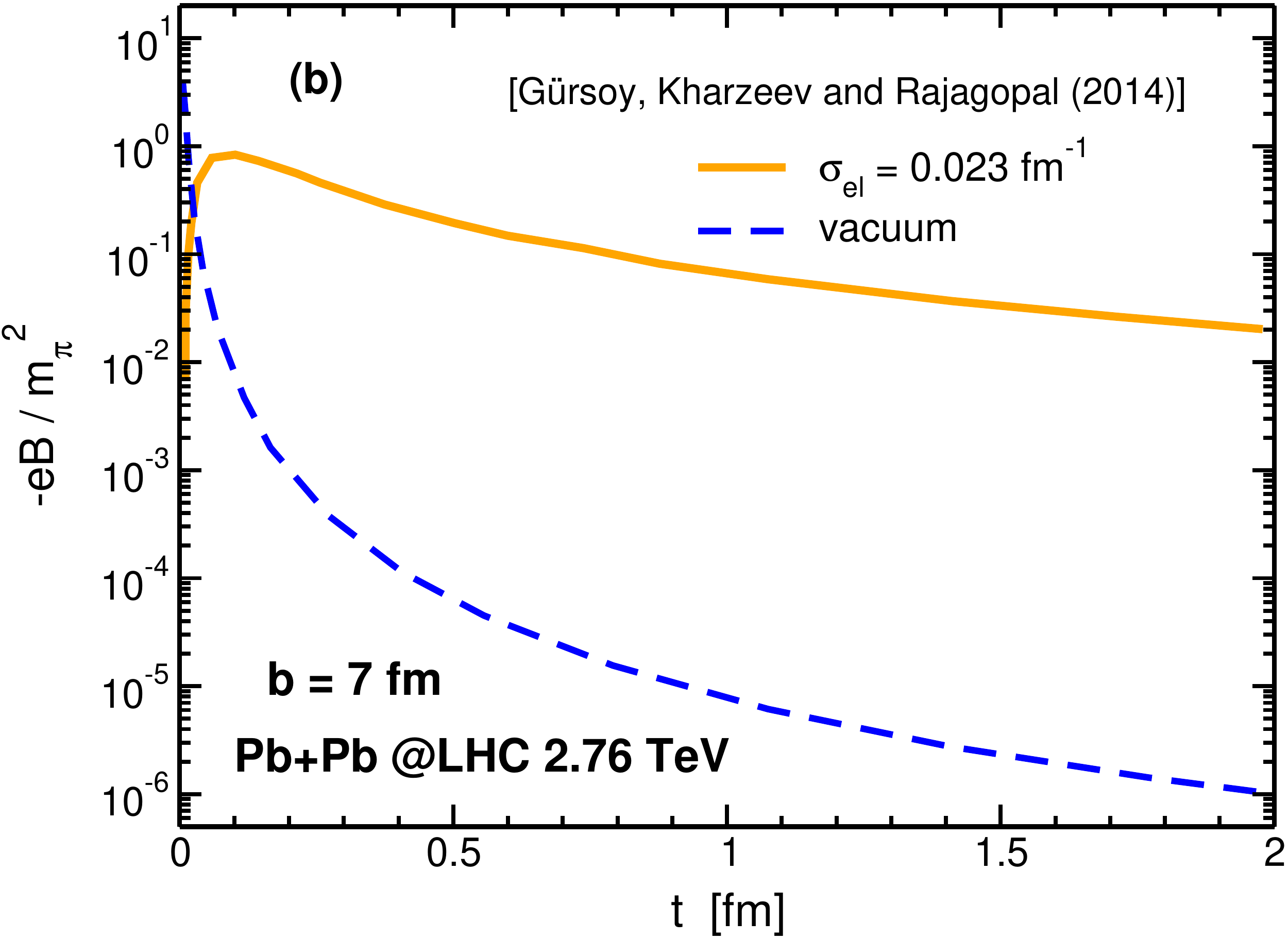}
\caption{(Color online) Temporal evolution of the magnetic field in the center of the overlapping region for symmetric heavy-ion collisions. Panel (a): results from Ref.~\cite{Voronyuk:2011jd} obtained with HSD simulations considering the fields produced by all charged particles (solid red line) or only spectator protons (dashed turquoise lines). Panel (b): results from Ref.~\cite{Gursoy:2014aka} obtained by means of semi-analytical computations of the field in the vacuum (dashed blue line) or in a conducting medium with electric conductivity $\sigma_{el}=0.023$ fm$^{-1}$ (solid orange curve).}
\label{fig:B_time_avg}
\end{figure*}

In Fig.~\ref{fig:B_time_avg} we represent the time evolution of the magnetic field in the center of the overlap area $(x,y,z)=(0,0,0)$ of symmetric HICs at ultrarelativistic energy obtained by means of different frameworks introduced in the Sec.~\ref{sec:emf_approaches}\footnote{In the original publications different conventions for units and colliding frames are used. Here the lines are uniformed to dimensionless unit and the reference frame convention used throughout this paper: the nucleus with centre at $x>0$ ($x<0$) moves with positive (negative) rapidity; therefore, the magnetic field points dominantly along the negative $y$ axis.}.
\\
In panel (a) we show the results obtained in Ref.~\cite{Voronyuk:2011jd} with the Hadron-String Dynamics (HSD) model for Au+Au collisions at $\sqrt{s_{NN}}=200$ GeV with impact parameter $b=10$ fm.
The solid red curve is the full computation of the magnetic field produced by all charges in the collisions, i.e. spectators, participants and newly produced particles; the dashed turquoise lines indicates the field generated by only spectator protons.
We see that in the first fm$/c$ after the collision of the two nuclei the magnetic field is mainly due to the spectator protons, drops down by three orders of magnitude and then become comparable with that from participant protons and newly-formed particles, which represent the dominant contribution at subsequent times.
The early-stage evolution is close to the field decay in the vacuum, as predicted firstly in Ref.~\cite{Kharzeev:2007jp}, with some differences: in Ref.~\cite{Kharzeev:2007jp} the colliding nuclei are treated as infinitely thin charged layers, neglecting their finite size in order to have a semi-analytical form of $B$, and the rapidity degradation of this pancake-shaped nuclei is simulated by means of a heuristic function, hence taking into account participant baryons but disregarding newly-created particles that are instead considered in the full computation (solid red line) with the HSD model.
Similar results for the early-stage evolution have been obtained in Ref.~\cite{Skokov:2009qp} through calculations within another microscopic transport model, i.e. the Ultrarelativistic Quantum Molecular Dynamics (UrQMD) approach, even though the latter includes only the spectator contribution and neglects the back reaction on the field of particle propagation, which are instead accounted for in HSD.
The HSD model is the precursor of PHSD \cite{Cassing:2009vt} introduced in the previous section, therefore the main results about the generation and evolution of the EMF continue to be valid within PHSD with some difference due to the fact that the partonic stage and the QCD phase transition between hadronic and deconfined matter are explicitely taken into account in PHSD, so that the contribution of the QGP is included in the computation of the EMF.
\\
In Fig.~\ref{fig:B_time_avg}(b) we plot the temporal evolution of $B_y$ in Pb+Pb collisions at $\sqrt{s_{NN}}=2.76$ TeV with $b=7$ fm obtained in Ref.~\cite{Gursoy:2014aka} through numerical calculations based on analytical solutions of the Maxwell equations for different values of the electric conductivity, i.e. solving Eq.~\eqref{eq:wave}. The solid orange curve representing the $B_y$ evolution in a plasma with electric conductivity $\sigma_{el}=0.023$ fm$^{-1}$ is compared with the magnetic field decay in the vacuum (i.e., $\sigma_{el}=0$ fm$^{-1}$) which is indicated by the dashed blue line and correspond to the calculations of Ref.~\cite{Kharzeev:2007jp}.
We see that the presence of a conducting plasma with nonzero conductivity strongly delays the decay of the magnetic field, with a more flat time evolution which resembles the behaviour for $t\gtrsim 1$ fm$/c$ of $B$ produced by all charged particles in the HSD model (solid red curve in Fig.~\ref{fig:B_time_avg}(a)).
However, in Ref.~\cite{Gursoy:2014aka} $B(t)$ has a mild decay since the very early times.
This is due to the simplification used in this approach, as well as in previous similar computations obtained in Refs.~\cite{Tuchin:2013ie,Tuchin:2013apa}, to treat the electrical conductivity as a constant in order to perform an analytic calculation. This consists in dealing with EMF that evolves all the time in a conducting medium.
In a more realistic picture, the electric conductivity is temperature dependent, as shown in Fig.~\ref{fig:sigmael_T}, therefore it depends on spacetime points in the plasma and decreases during the plasma expansion and cooling. Moreover, it is not trivial its behaviour in the early stage before the onset of a mostly thermalized QGP, where $\sigma$ can not be described by its equilibrium value determined by lattice QCD calculations.
Hence, in the pre-equilibrium era the blue curve better approximates the $B$ evolution in the collision before the plasma formation.

\begin{figure*}[hbt!]
\centering
\includegraphics*[trim={35 5 5 15},clip,width=0.65\columnwidth]{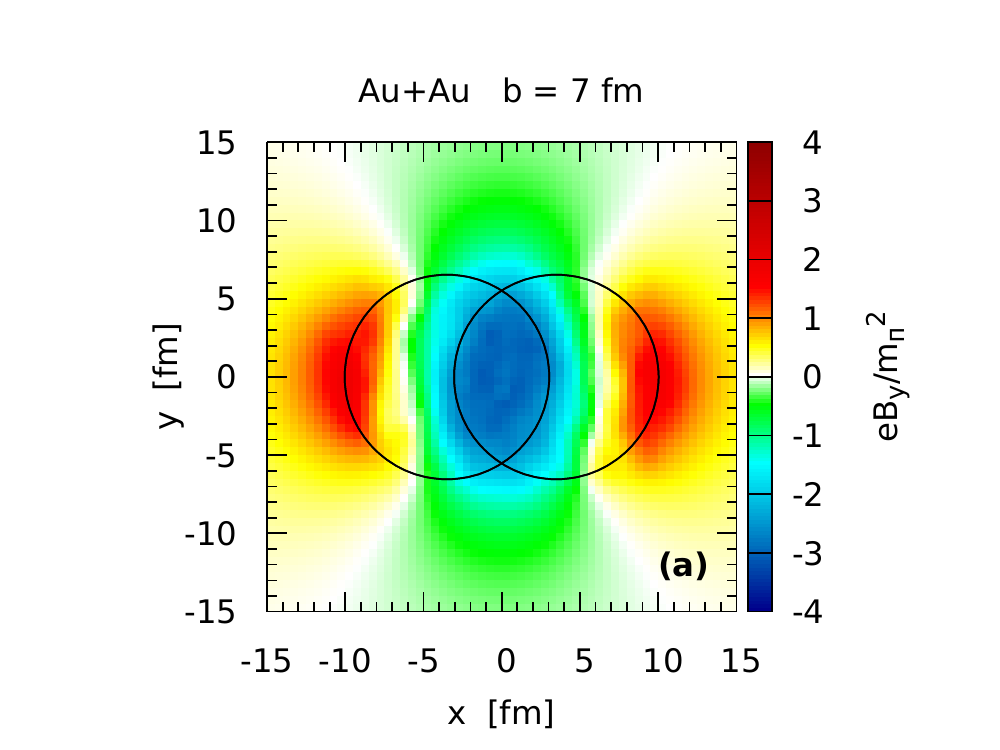}
\includegraphics*[trim={35 5 5 15},clip,width=0.65\columnwidth]{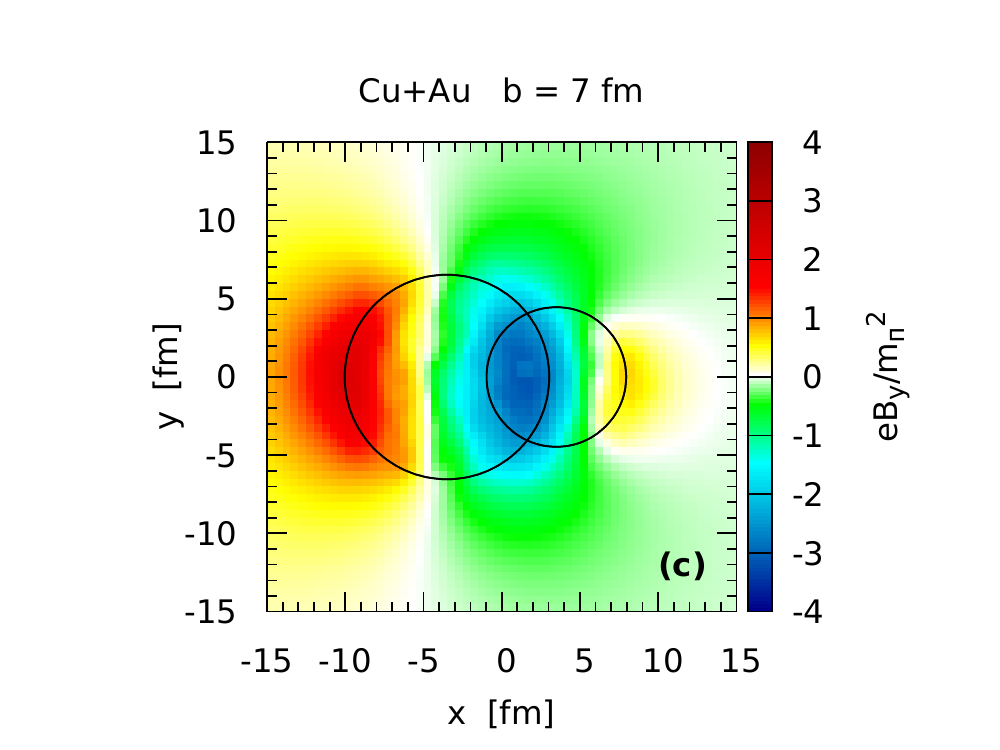}
\includegraphics*[trim={35 5 5 15},clip,width=0.65\columnwidth]{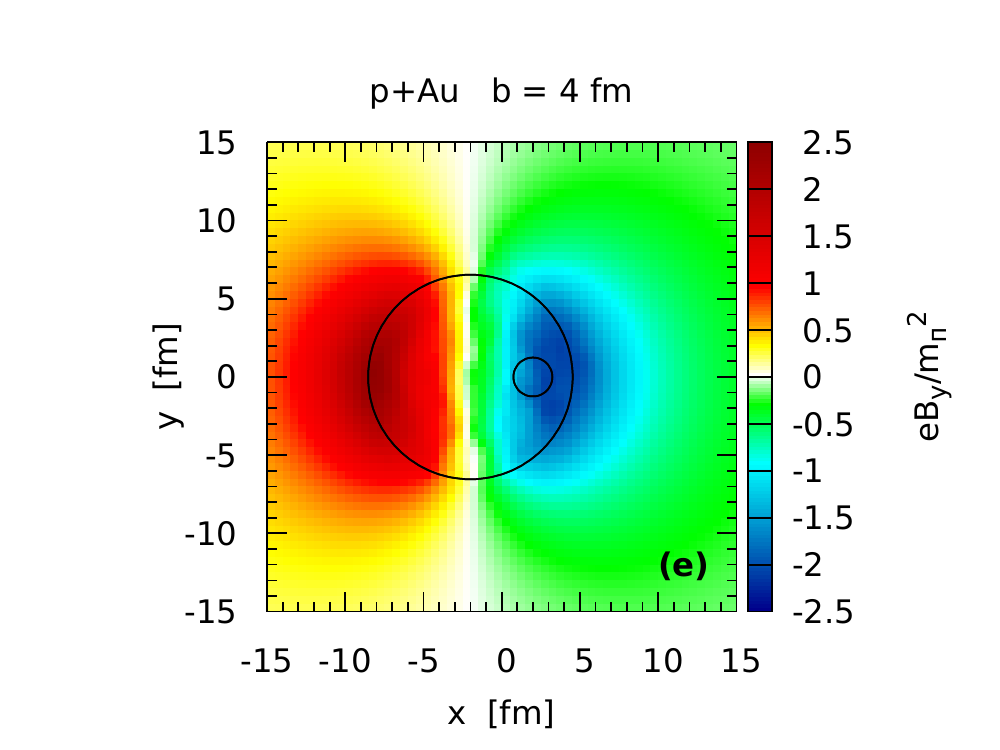}
\\[\baselineskip]
\includegraphics*[trim={35 5 5 15},clip,width=0.65\columnwidth]{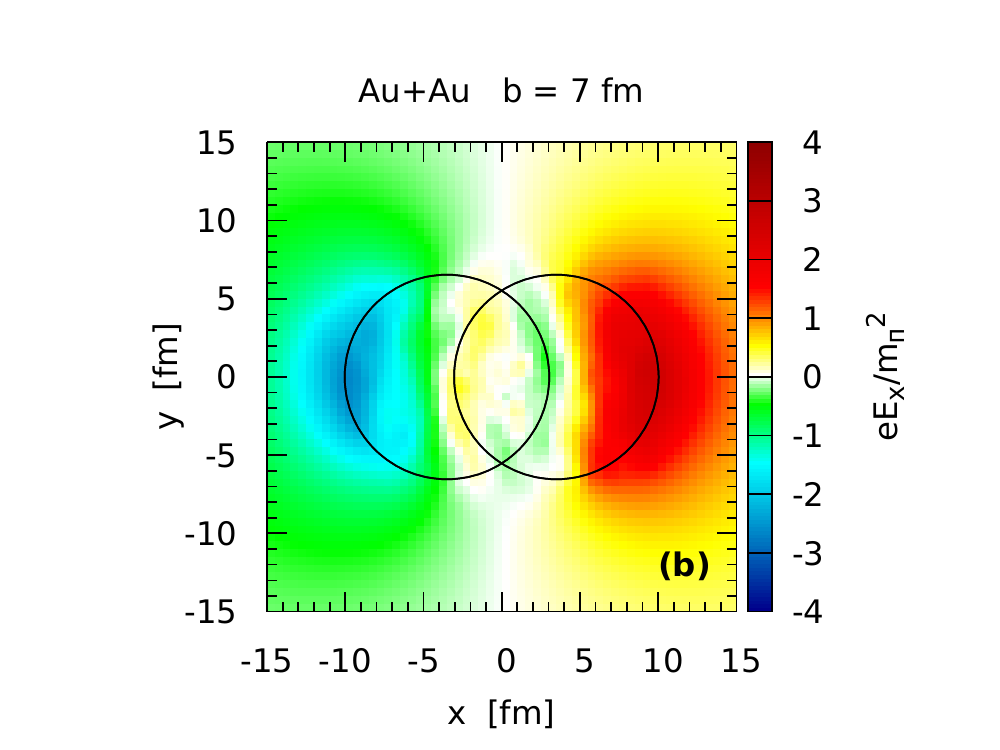}
\includegraphics*[trim={35 5 5 15},clip,width=0.65\columnwidth]{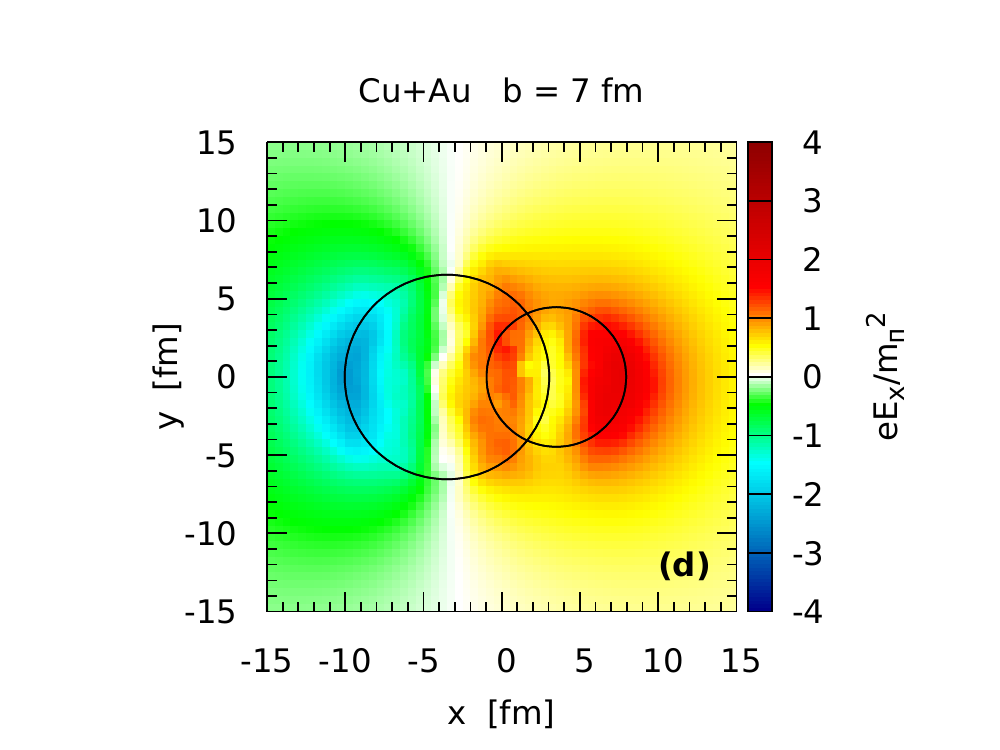}
\includegraphics*[trim={35 5 5 15},clip,width=0.65\columnwidth]{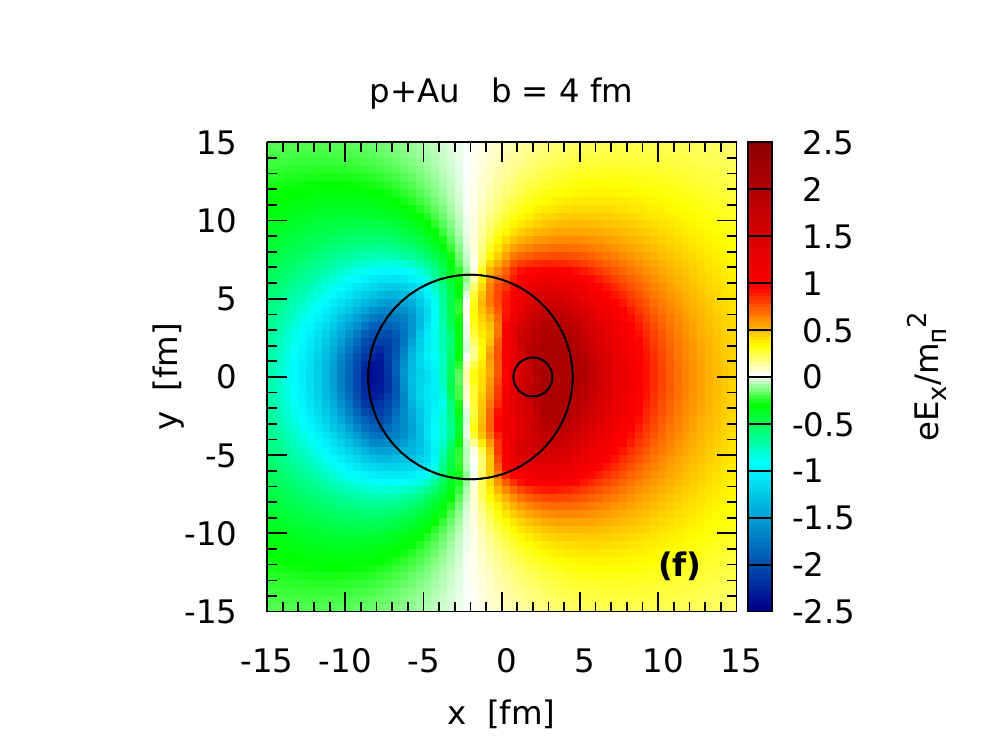}
\caption{(Color online) Transverse profile of the electromagnetic field components $B_y$ and $E_x$ in $z=0$ at the maximum overlap time of the two nuclei for Au+Au collisions with impact parameter $b=7$ fm (a--b), Cu+Au collisions with $b=7$ fm (c--d) and p+Au collisions with $b=4$ fm (e--f) at $\sqrt{s_{NN}}=200$ GeV. The calculations are performed by means of the PHSD model with the inclusion of the electromagnetic fields \cite{Voronyuk:2011jd,Voronyuk:2014rna,Oliva:2019kin}. Circles representing the position of the colliding nuclei are drawn to guide the eye.}
\label{fig:EMF_space}
\end{figure*}

Up to now we have considered the time evolution of the event-averaged EMF in symmetric collisions, which is dominated by the magnetic field $B_y$ and the Faraday-induced electric field $E_x$, whereas the others components are nearly vanishing.
In asymmetric systems the difference in the proton number of the initial nuclei generates a substantial Coulomb electric field $E_x$ directed from the heavier nucleus towards the lighter one.
Proton-induced collisions represent the most extreme case in which the EMF are basically produced by the heavy nucleus, leading to very similar values for the magnetic field $B_y$ and the electric field $E_x$.

Among the various theoretical models addressing the topic of the EMF generated in non-central collisions, the PHSD approach described in Sec.~\ref{sec:emf_approaches} has been used to perform detailed calculations of the EMF and their observable effects in all three cases of symmetric \cite{Voronyuk:2011jd,Toneev:2011aa,Toneev:2012zx}, asymmetric \cite{Voronyuk:2014rna,Toneev:2016bri} and small systems \cite{Oliva:2019kin}.
In Fig.~\ref{fig:EMF_space} we show the PHSD calculations of the $B_y$ (upper panels) and $E_x$ (lower panels) components of the EMF produced at top RHIC energy in Au+Au collisions at $b=7$ fm (a--b), Cu+Au at $b=7$ fm (c--d) and p+Au collisions at $b=4$ fm (e--f).
The field strength are computed at $z=0$ and at the maximum overlapping time of the collision, namely when the centres of the two colliding nuclei lie in the same transverse plane.
The circles drawn in all panels help to roughly identify the size and position of the nuclei in the three collision systems and to highlight the interaction area.
We see that in Au+Au collisions the dominant $B_y$ component (a) reaches value $\left|eB_y\right|\simeq 4\,m_{\pi}^2$ whereas $E_x$ (b) is nearly vanishing in the intersection area, as well as all other EMF components not shown in the figure.
\\
In the case of Cu+Au collisions, besides a huge magnetic field $B_y$ (c), also the electric field $E_x$ (d) reaches comparable values of the order of $m_{\pi}^2$. It is directed from the heavier gold nucleus towards the lighter copper nucleus, hence the overlapping area is interested by an intense and asymmetrically distributed electric field. 
\\
The initial EMF produced in p+Au collisions correspond basically to that produced by the gold nucleus moving at ultrarelativistic velocity and, consequently, it does not show a significant dependence on the impact parameter of the collision. However, the impinging proton and the particles created after the collision feel a different strength of the EMF depending on the collision point.
The $E_x$ field (f) is strongly asymmetric inside the overlap area of non-central p+Au collisions and has a magnitude similar to that of the $B_y$ component (e), both reaching values of about $\simeq 2\,m_{\pi}^2$ in collisions at $b=4$ fm; the other EMF components are almost vanishing.

We have discussed the spacetime profiles of the EMF averaged over many events (Fig.~\ref{fig:B_time_avg}(a), Fig.~\ref{fig:EMF_space}) or obtained with event-averaged charge density (Fig.~\ref{fig:B_time_avg}(b)).

However, event-by-event fluctuations in proton positions in the colliding nuclei lead to event-by-event fluctuations of the generated EMF \cite{Bzdak:2011yy,Deng:2012pc,Toneev:2012zx}.
This means that, even though on average the only
nonvanishing component of the field are $B_y$ and $E_x$, in a single event other components of the EMF can reach very high and comparable values,
\begin{equation}
\langle\vert B_x \vert\rangle \approx \langle\vert B_y \vert\rangle \approx \langle\vert E_x \vert\rangle \approx \langle\vert E_y \vert\rangle.
\end{equation}
This happens even in the central point of the overlap area in symmetric collisions, where from symmetry considerations one expect that on average the only nonvanishing component of the field is $B_y$, and also for very central collisions, where even $B_y$ is expected to disappear.
Moreover, $\langle\vert B_x \vert\rangle$, $\langle\vert E_x \vert\rangle$ and $\langle\vert E_y \vert\rangle$ show a very mild dependence on the impact parameter up to $b\sim R_A$, being $R_A$ the radius of the nucleus.

The estimates of the EMF strengths provide limited information on the influence of the fields on particle propagation and final observables. A more clear picture arises by looking at the total momentum increment $\Delta{\bm p}$ obtained by summing over subsequent time steps the mean increase of charged-particle momenta due to the action of the electric and magnetic forces during the short time interval \cite{Toneev:2011aa,Toneev:2012zx}.
For all components of $\Delta{\bm p}$ the contributions from the electric and magnetic fields in the Lorentz force \eqref{eq:lorentz} largely compensate each other, with a tiny unbalance of the two terms especially in the momentum increment along the $x$ direction, as clearly shown for Au+Au collision at top RHIC energy in Refs.~\cite{Toneev:2011aa,Toneev:2012zx} by means of PHSD microscopic calculations.
This compensation effect can be easily illustrated in the simplified one-dimensional case for a particle with charge $e$ at position $x=x(t)$:
$$
eE\sim-e\dfrac{\partial A}{\partial t}= -e\dfrac{\partial A}{\partial x}\dfrac{\partial x}{\partial t}\sim -eBv,
$$
namely the forces due to the electric and magnetic transverse components are roughly equal and inversely directed \cite{Toneev:2011aa,Toneev:2012zx}.
Thus, the influence of the EMF in relativistic collisions can not be reliably studied neglecting the electric field; moreover, the impact on final observables is expected to be small not only for the shortness of the interaction time due to a fast decay of the fields but also for the partial cancellation of the forces determined by the transverse electric and magnetic fields.

\section{Directed flow in symmetric systems}
\label{sec:v1_sym}

The rapidity-odd $v_1$ is sensitive to the vortical structure generated in the fireball because of the intense angular momentum of the two colliding nuclei.
In order to extract the imprint of the EMF one has to study the charge-odd directed flow, by looking at particle and antiparticle of a specific species separately or considering positively and negatively charged hadrons.
The intense EMF produced by the moving charges of spectators and newly produced particles generate a sidewards deflection of quark and antiquarks in opposite direction according to their electric charge, ending up with a different $v_1$ of the final particles in which they hadronize.

\begin{figure*}[t!]
\centering
\includegraphics[trim={0 0 0 0},clip,width=0.65\columnwidth]{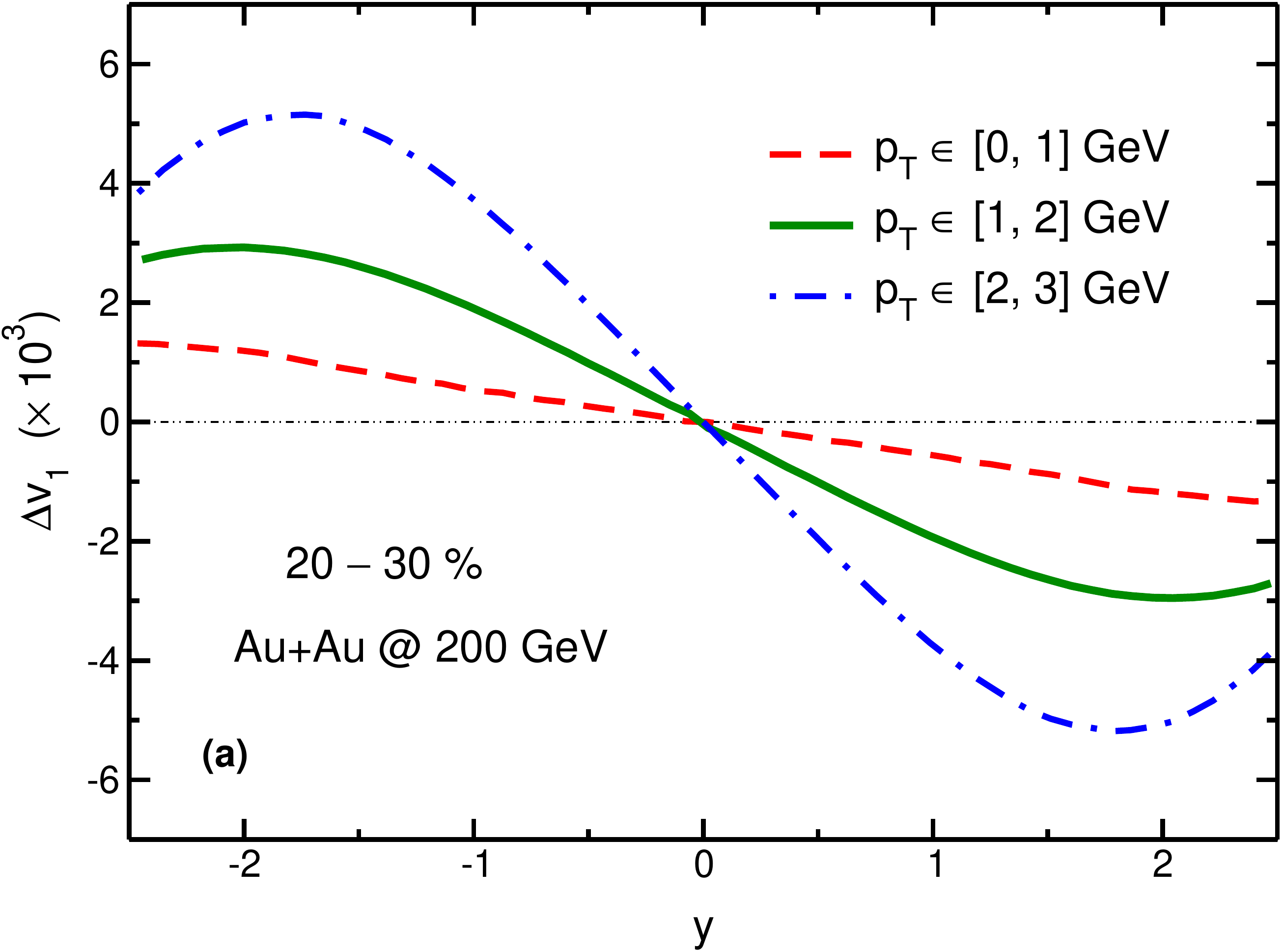}\quad
\includegraphics[trim={0 0 0 0},clip,width=0.65\columnwidth]{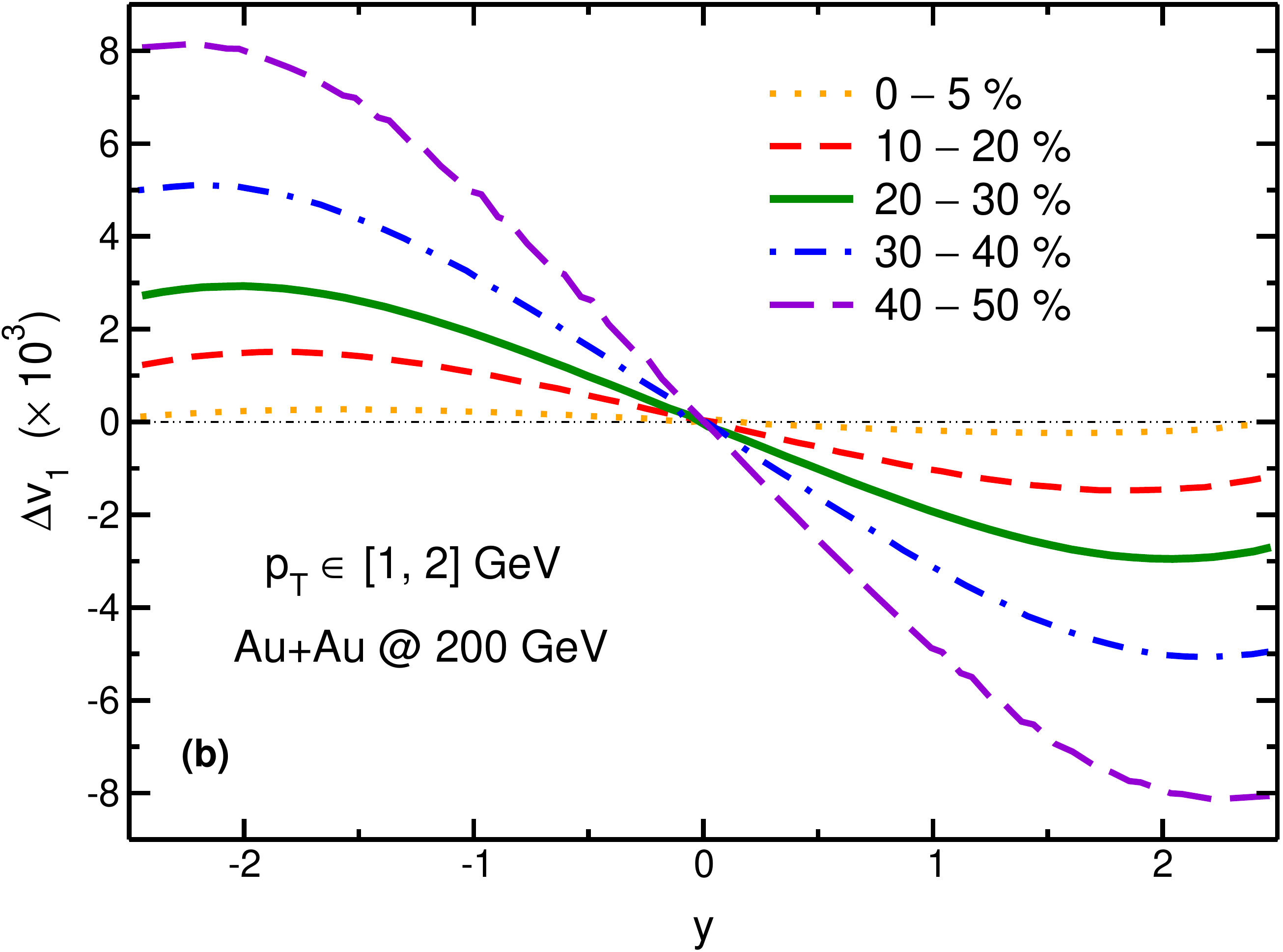}\quad
\includegraphics[trim={0 0 0 0},clip,width=0.65\columnwidth]{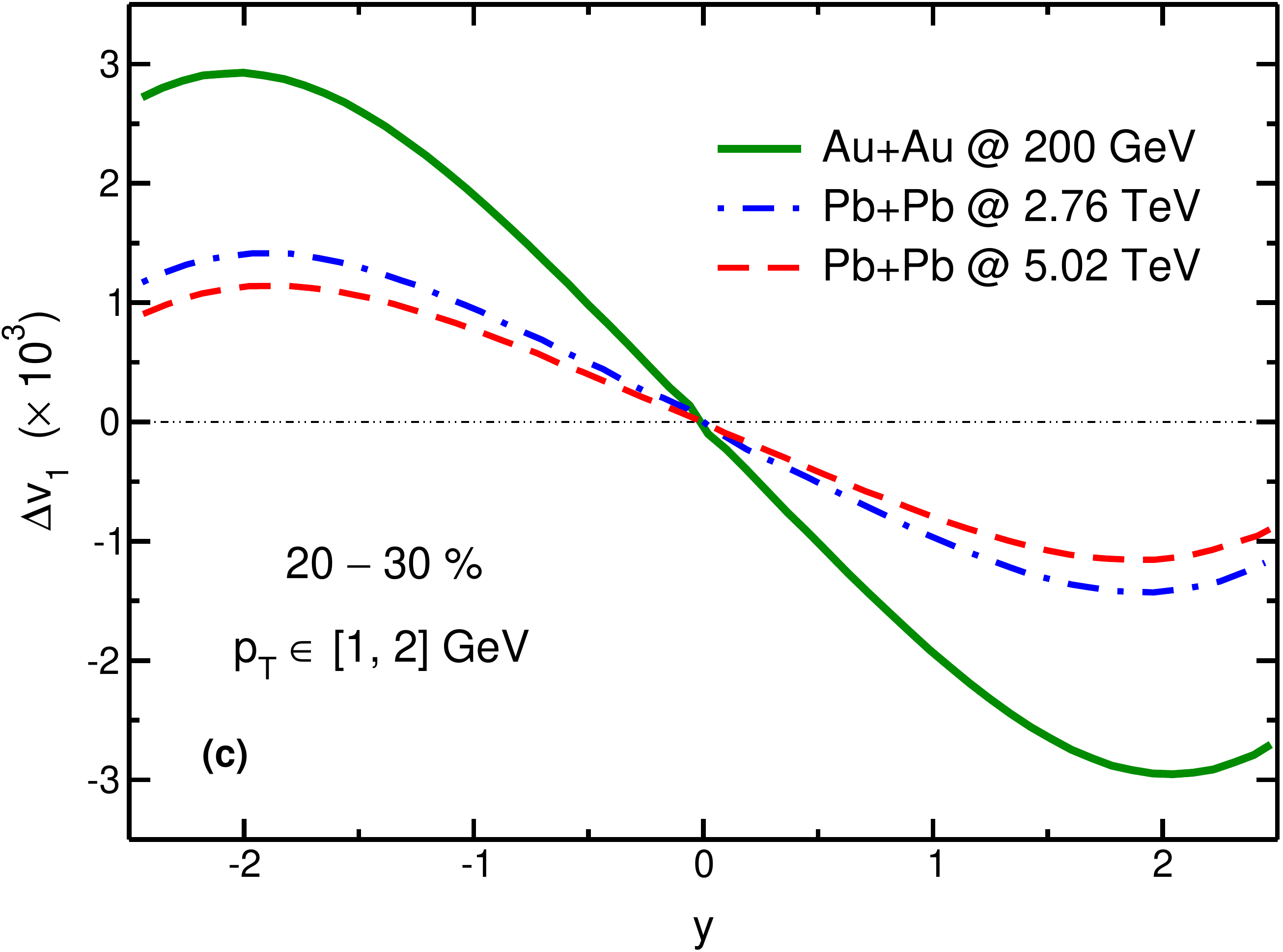}
\caption{(Color online) Rapidity dependence of the electromagnetically-induced splitting of the directed flow of pions $\Delta v_1^{\pi}=v_1(\pi^+)-v_1(\pi^-)$ in the transverse momentum range $1<p_T<2$ GeV for $20-30\%$ Au+Au collisions at $\sqrt{s_{NN}}=200$ GeV (solid green line) in comparison with the $v_1$ in different transverse momentum ranges (a), centrality classes (b) and collision energies (c). The results are from Ref.~\cite{Gursoy:2018yai} and are obtained with hydrodynamical calculations within the iEBE-VISHNU framework taking into account the electromagnetic fields.}
\label{fig:v1_y_AuAu200GeV_pions}
\end{figure*}

The directed flow of light hadrons in ultrarelativistic HICs has been studied and discussed in many works; in particular, the effect of the EMF has been highlighted in Refs.~\cite{Gursoy:2014aka,Gursoy:2018yai} within the hydrodynamic framework, mainly focusing on the rapidity dependence of the $v_1$ of charged pions, protons and antiprotons.
\\
In Ref.~\cite{Gursoy:2018yai} the QGP evolution is described by means of the iEBE-VISHNU approach \cite{Shen:2014vra} which evolves the equations of relativistic viscous hydrodynamic assuming longitudinal boost-invariance.
On top of this hydrodynamical background the Maxwell equations are solved through Eq.~\eqref{eq:wave} for describing the evolution of the EMF generated by electric charges and currents due to spectator protons and evolving in a plasma with constant electrical conductivity, as explained in Sec.~\ref{sec:emf_approaches}.
\\
From the momentum distribution of particles with different charge coming out from the calculation, the splitting between the directed flow for positively and negatively charged particles is evaluated by the difference
\begin{equation}
\Delta v_1 \equiv v_1(+) - v_1(-),
\label{eq:v1diff}
\end{equation}
in order to isolate the $v_1$ induced by the EMF from the much larger contribution due to the background hydrodynamic flow.
\\
Some of the results of the calculations of Ref.~\cite{Gursoy:2018yai} for the $v_1$ splitting of $\pi^+$ and $\pi^-$ are shown in Fig.~\ref{fig:v1_y_AuAu200GeV_pions}.
In all panels the solid green line correspond to $\Delta v_1$ of pions in the transverse momentum range $1<p_T<2$ GeV for $20-30\%$ Au+Au collisions at $\sqrt{s_{NN}}=200$ GeV.
Panel (a) represents the comparison with different transverse momentum integration range: the $\Delta v_1$ clearly increases as the $p_T$ values increase.
The authors found the the same trend holds for the $\Delta v_1$ of protons--antiprotons.
In panel (b) the centrality dependence is depicted, showing that the $\Delta v_1$ of pions increases going from central towards peripheral collisions. This is in line with the fact highlighted also in other works \cite{Voronyuk:2011jd,Gursoy:2014aka} that the effect of EMF on final observables, in particular on the directed flow, is mainly driven by the fields produced by spectator protons with respect to the smaller contribution generated by the participants; indeed, more peripheral collisions correspond to a higher number of spectators.
The electromagnetically-induced $v_1$ splitting of pions decreases as the collision energy increases, as shown in the right panel of Fig.~\ref{fig:v1_y_AuAu200GeV_pions}, where the result at top RHIC energy is compared with Pb+Pb collisions at $\sqrt{s_{NN}}=2.76$ TeV (dot-dashed blue line) and $\sqrt{s_{NN}}=5.02$ TeV (dashed red line).
The reason is that in collisions at higher energy the spectators move away more quickly from the centre of the collision, hence the EMF generated by them (which is the main contribution) experience a faster decrease with time; as a consequence their effect is milder than at lower energy collisions.
The same qualitative behaviour according to the charge of the particle has been shown in Ref.~\cite{Gursoy:2018yai} for the difference in the directed flow of protons and antiprotons $\Delta v_1^p\equiv v_1(p)-v_1(\overline{p})$: at forward rapidity $v_1(p)<v_1(\overline{p})$ and at backward rapidity $v_1(p)>v_1(\overline{p})$.
\\
Hence, at $y>0$ positively charged particles ($p$, $\pi^+$) are pushed downward and negatively charged particles ($\overline{p}$, $\pi^-$) are pushed upward along the impact parameter direction by the Lorentz force due to the EMF; this small splitting comes out from the almost complete local cancellation of the two opposite forces due to the electric part and the magnetic part of the Lorentz force and indicates that the total effect is mildly dominated by the electric field. 
\\
The results of Ref.~\cite{Gursoy:2018yai} shown in Fig.~\ref{fig:v1_y_AuAu200GeV_pions} give an overview of the theoretical expectations for the rapidity dependence of the electromagnetic splitting of the $v_1$ of charged pions as a function of the transverse momentum interval, the centrality class and the beam energy of the collision.
Pions are somehow a more clean probe of the EMF with respects to other light hadrons in which there are contributions of $\Delta v_1$ coming from other sources. However, the final effect is tiny and requires very precise experimental measurements.

Besides light hadrons, very promising and interesting probes of the initial EMF are the heavy mesons, in particular the neutral $D^0$ and $\overline{D}^0$, as predicted in Ref.~\cite{Das:2016cwd} by means of the Langevin dynamics for charm and anticharm quarks in an expanding QGP background. The initial conditions for solving the relativistic Langevin equation for the heavy quarks and the relativistic transport code for the bulk medium are constrained by the experimental data on the nuclear modification factor $R_{AA}(p_T)$ and the elliptic flow $v_2$ of $D$ mesons \cite{Das:2015ana} and the transverse momentum spectra and the $v_2$ of the bulk \cite{Greco:2008fs,Ruggieri:2013ova}.
The authors of Ref.~\cite{Das:2016cwd} show that the effect of the EMF on the $v_1$ of charm quarks is significantly larger than that of light quarks because the heavy quarks, being their formation time scale of the order of 0.1 fm$/c$ (much smaller than that of light quarks), are already present when the EMF reach their maximal magnitude. Moreover, the relaxation time of charmed particles is comparable to the QGP lifetime, thus helping them to retain the initial acceleration in the transverse direction caused by the electromagnetic field.
Hence, $D^0$ and $\overline{D}^0$, in spite of being neutral, are splitted by the EMF due to the electric charges of their heavy constituents. This can also be considered a further evidence of the presence of a deconfined phase in relativistic HICs.

In recent years the experimental results for the directed flow of charged hadrons as well as of identified light and heavy mesons has been reported by the STAR \cite{Adam:2019wnk} and ALICE Collaborations. The recent efforts have been focused to measure not only the combined $v_1$ of particle--antiparticle but also its difference, in order to extract the influence of the EMF through the charge-dependence effect that involves the final hadrons and their production channels.
In particular, we will discuss the experimental data for the directed flow of neutral $D$ mesons in comparison to the theoretical expectations and to the measurements for light hadrons. 

\begin{figure}[t!]
\centering
\includegraphics[trim={0 0 0 0},clip,width=0.9\columnwidth]{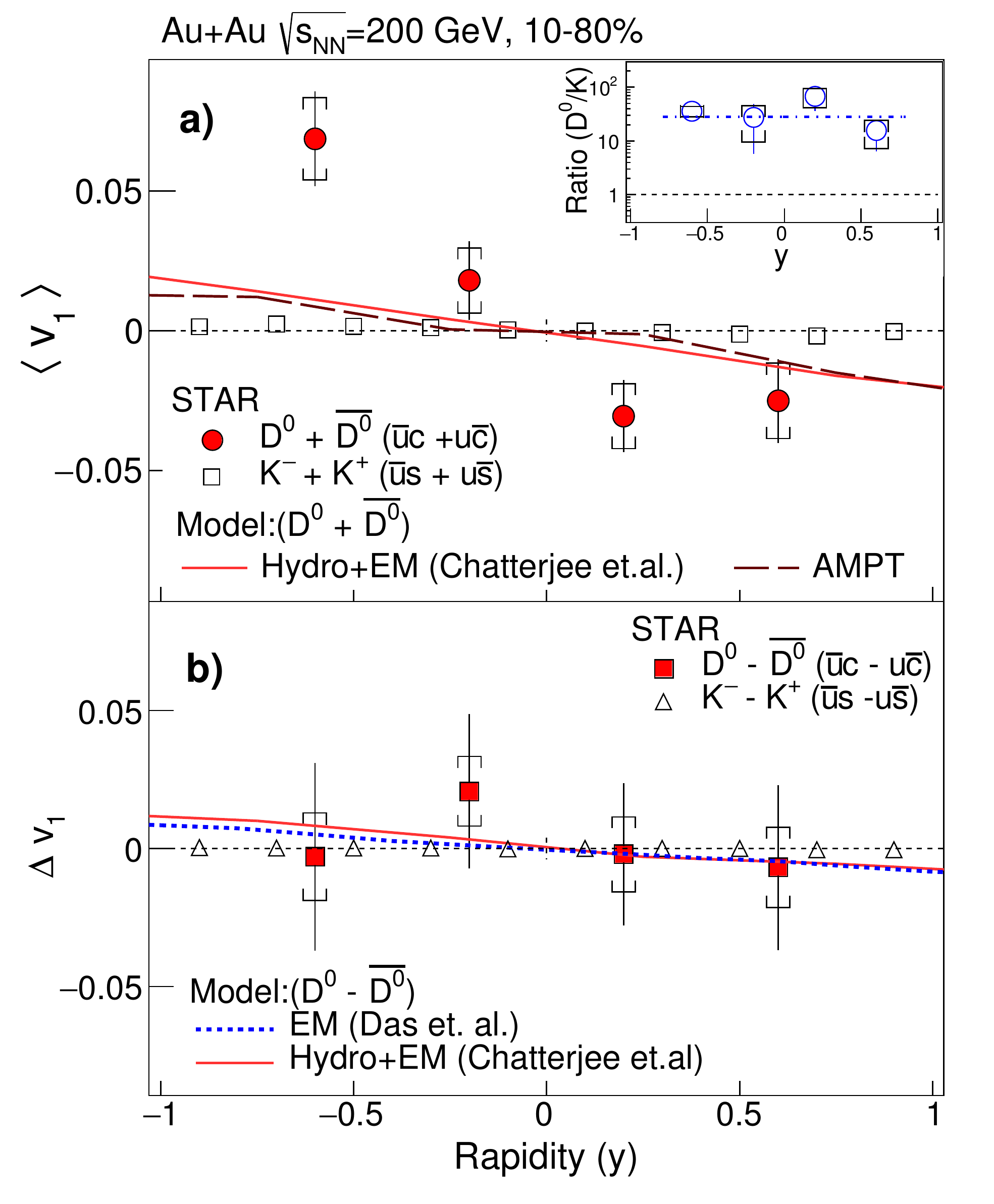}
\caption{(Color online) Directed flow average $\langle v_1\rangle$ (a) and difference $\Delta v_1$ (b) of heavy and light mesons as a function of rapidity for 10--80\% Au+Au collisions at $\sqrt{s_{\rm NN}}=200$ GeV measured by the STAR Collaboration \cite{Adam:2019wnk} in comparison to the theoretical calculations. Filled red symbols represent the experimental data for $D^{0}$ and $\overline{D^{0}}$ mesons in the transverse momentum range $p_T>1.5$ GeV; the measurements for the $v_1$ of $K^+$ and $K^-$ integrated for $p_T>0.2$ GeV are labelled by empty markers \cite{Adamczyk:2017nxg}. The inset shows the ratio between the $v_{1}$ of neutral $D$ mesons and that of charged kaons. The error bars and caps represent, respectively, statistical and systematic uncertainties. In panel (a) the solid red line correspond to hydrodynamic simulations with the inclusion of the electromagnetic fields (``Hydro+EM'') from Refs.~\cite{Chatterjee:2017ahy,Chatterjee:2018lsx} and the dashed brown line is the results of transport calculations with the AMPT model \cite{Nasim:2018hyw}. In panel (b) the dotted blue line denotes the prediction reported in Ref.~\cite{Das:2016cwd} with a Langevin approach coupled to the electromagnetic fields (``EM'') and the solid red line is the calculation within the ``Hydro+EM'' framework \cite{Chatterjee:2017ahy,Chatterjee:2018lsx}. The figure is taken from Ref.~\cite{Adam:2019wnk}.}
\label{fig:v1_y_RHIC}
\end{figure}

Fig.~\ref{fig:v1_y_RHIC} shows the STAR data in  10--80\% Au+Au collisions at RHIC energy of $\sqrt{s_{\rm NN}}=200$ GeV \cite{Adam:2019wnk} along with predictions from theoretical models \cite{Das:2016cwd,Chatterjee:2017ahy,Chatterjee:2018lsx,Nasim:2018hyw}.
In panel (a) the rapidity dependence of the averaged directed flow $\langle v_1 \rangle$ of neutral $D$ mesons for $p_T>1.5$ GeV measured by STAR is denoted by red circles. They are compared to the measurements for combined directed flow of charged kaons with transverse momentum $p_T>0.2$ GeV represented by open squares \cite{Adamczyk:2017nxg}.
The strength of the $v_1$ is often identified by its slope $dv_1/dy$ at midrapidity.
By fitting the $v_1(y)$ of the averaged $D^{0}$ and $\overline{D^{0}}$ with a linear function constrained to pass through the origin, the STAR Collaboration found the fit
$$
d\langle v_1\rangle^D/dy=-0.080\pm 0.017\,\rm{(stat.)}\pm 0.016\,\rm{(syst.)};
$$
the fit for the slope of charged kaons is
$$
d\langle v_1\rangle^K/dy=-0.0030\pm 0.0001\,\rm{(stat.)}\pm 0.0002\,\rm{(syst.)}.
$$
Therefore, the STAR Collaboration observed an absolute value of the neutral $D$ mesons $dv_1/dy$ that is about 25 times larger than that of the charged kaons with a $3.4\sigma$ significance \cite{Adam:2019wnk} and is the largest among that of all the eleven particle species measured at top RHIC energy \cite{Adamczyk:2011aa,Adamczyk:2014ipa,Adamczyk:2017nxg}.
This supports the idea that heavy mesons are excellent probes to investigate the early dynamics of nuclear collisions.
\\
The ``antiflow'' behaviour -- the negative slope at midrapidity -- of the directed flow of $D$ mesons at top RHIC energy has been predicted fifteen years ago in transport calculations within the Hadron-String Dynamics (HSD) approach \cite{Bratkovskaya:2004ec}. The authors of Ref.~\cite{Chatterjee:2017ahy,Chatterjee:2018lsx}, by means of a Langevin dynamics for heavy quarks coupled to hydrodynamic equations for the bulk medium, have highlighted for the first time that the $v_1$ of $D$ mesons in noncentral collisions is several times larger than that of charged particles; their prediction (solid red line labelled as ``Hydro+EM'') is shown in Fig.~\ref{fig:v1_y_RHIC}(a) along with the result from A Multi-Phase Transport (AMPT) model \cite{Nasim:2018hyw} (dashed brown curve).
Even though the qualitative behaviour of the $D$-meson $v_1$ is captured by the theoretical calculations, the experimental data are quantitatively underestimated. 

In Fig.~\ref{fig:v1_y_RHIC}(b) the experimental data for the directed flow splitting of neutral $D$ mesons $\Delta v_1^D=v_1(D^{0})-v_1(\overline{D^{0}})$ measured by the STAR Collaboration \cite{Adam:2019wnk} is shown with red squares together with the theoretical predictions from Ref.~\cite{Das:2016cwd} labelled by the dotted blue line (``EM'') and from Ref.~\cite{Chatterjee:2018lsx} represented by the solid red line (``Hydro+EM'').
The authors of Ref.~\cite{Das:2016cwd}, by solving the Langevin equation in an expanding QGP background described with a relativistic Boltzmann approach, have predicted that the $v_1$ splitting of $D^{0}$ and $\overline{D^{0}}$ mesons is orders of magnitude larger than that of light hadron species, since the heavy $c$ and $\overline{c}$ quarks are produced in hard scatterings and hence feel the high early EMF.
A similar result has been obtained within the Langevin+hydrodynamics approach of Ref.~\cite{Chatterjee:2017ahy} with the inclusion of the EMF \cite{Chatterjee:2018lsx}.
The linear fit of the $\Delta v_1$ slope (constrained to intersect the origin) provided by STAR is given by
$$
d\Delta v_1^D/dy=-0.011 \pm 0.034\,\rm{(stat.)}\pm 0.020\,\rm{(syst.)};
$$
within the present uncertainties the data are consistent also with a zero slope.
\\
The theoretical results of the $\Delta v_1$ of $D$ mesons are dependent on the assumed electric conductivity of the bulk medium which affect the time evolution of the EMF, as discussed in Sec.~\ref{sec:emf}, and on the description of the dynamics of charm and anticharm quarks in the QGP.
Both models predicted a nonzero slope \cite{Das:2016cwd,Chatterjee:2018lsx} which is in the right ballpark considering the current experimental errors.

\begin{figure}[t!]
\centering
\includegraphics[trim={0 0 0 0},clip,width=0.95\columnwidth]{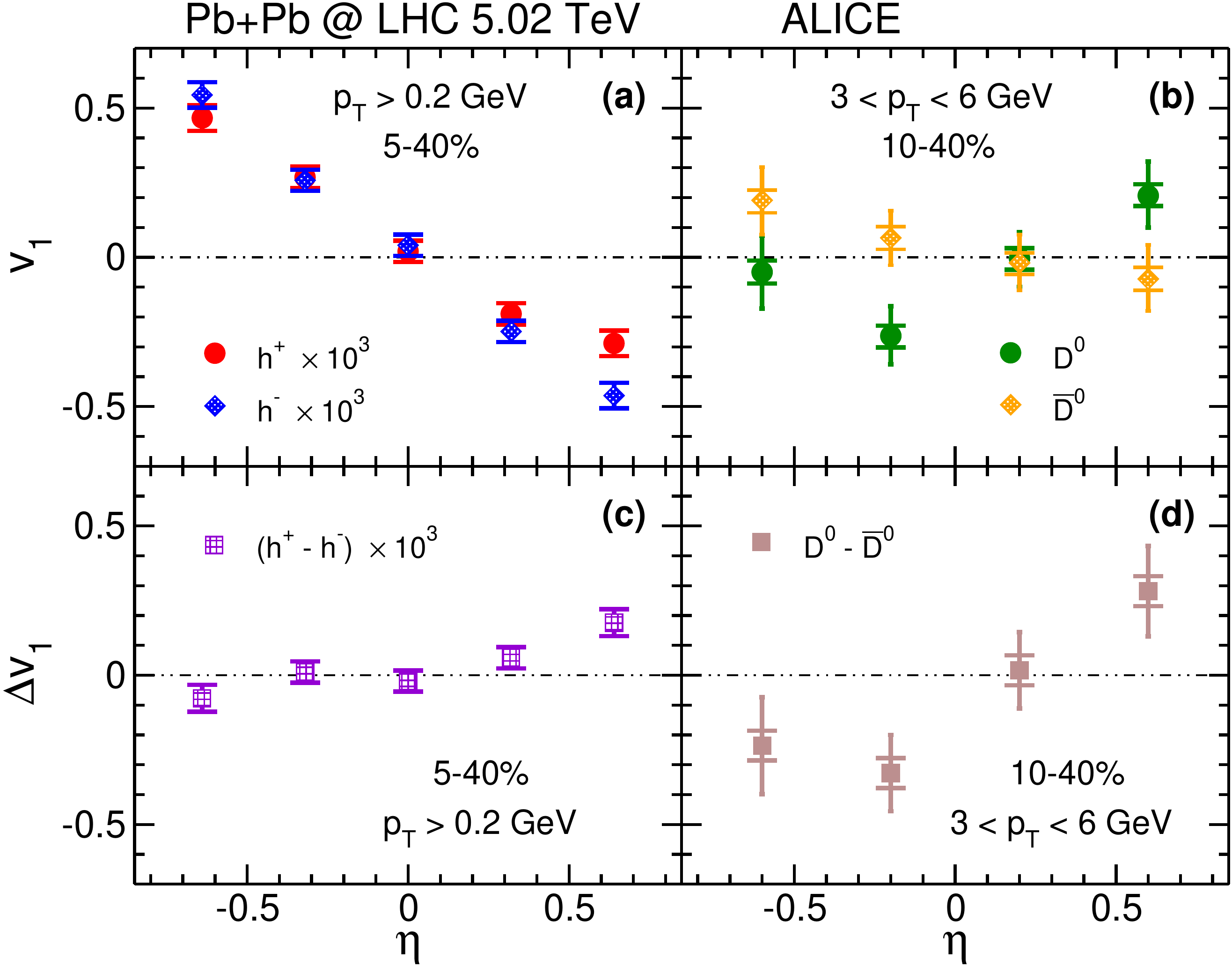}
\caption{(Color online) Pseudorapidity dependence of the directed flow and its splitting measured by ALICE Collaboration \cite{Acharya:2019ijj} for Pb+Pb collisions at $\sqrt{s_{\rm NN}}=5.02$ TeV.
Left panels: (a) $v_1$ of positively (red circles) and negatively (blue diamonds) charged hadrons and (c) its difference $\Delta v_1(h)=v_1(h^+)-v_1(h^-)$ (violet squares) in the 5--40\% centrality class integrated for transverse momenta $p_T>0.2$ GeV.
Right panels: (b) $v_1$ of $D^0$ (green circles) and $\overline{D}^0$ (orange diamonds) mesons and (d) its difference $\Delta v_1(D) = v_1(D^0) - v_1(\overline{D}^0)$ (brown squares) for 10--40\% central collisions and $3<p_T<6$ GeV. The results for charged hadrons are multiplied by 1000.}
\label{fig:v1_eta_LHC}
\end{figure}

The ALICE measurements of the directed flow $v_1$ of charged hadrons and heavy mesons as a function of pseudorapidity $\eta$ in Pb+Pb collisions at $\sqrt{s_{\rm NN}}=5.02$ TeV \cite{Acharya:2019ijj} are shown in Fig.~\ref{fig:v1_eta_LHC}.
\\
In panel (a) red circles and blue diamonds represent, respectively, the directed flow of positively and negatively charged hadrons integrated for transverse momenta $p_T>0.2$ GeV in 5--40\% central collisions.
As for collisions at RHIC energy, charged hadrons show the ``antiflow'' nature of the directed flow at midrapidity with similar magnitude for collisions at $\sqrt{s_{\rm NN}}=2.76$ TeV and $\sqrt{s_{\rm NN}}=5.02$ TeV.
The difference between the directed flow of positively and negatively charged hadrons $\Delta v_1^h=v_1(h^+)-v_1(h^-)$ is shown in panel (c) with violet squares.
The ALICE Collaboration extracted a fit of its slope with a linear function:
$$
d\Delta v_1^h/d\eta=\left[1.68\pm 0.49\,\rm{(stat.)}\pm0.41\,\rm{(syst.)}\right]\times 10^{-4},
$$
indicating a positive value with a significance of $2.6\sigma$.
\\
The theoretical prediction for the $v_1$ splitting between $\pi^+$ and $\pi^-$ at LHC energy obtained in Ref.~\cite{Gursoy:2018yai} with hydrodynamic simulations coupled to the EMF and shown in Fig.~\ref{fig:v1_y_AuAu200GeV_pions}(c) gives a $\Delta v_1$ of the same order of magnitude of that between positively and negatively charged hadrons measured by ALICE but with the opposite sign.
However, those calculations are able to capture the effect of the EMF on the $v_1$ but do not account for another source of $v_1$ that is the vorticity and the angular momentum of the colliding system and the baryon stopping mechanism, which act in a different way on $K^+$ and $p$ from one hand and on $K^-$ and $\overline{p}$ from the other hand \cite{Adamczyk:2014ipa,Adamczyk:2017nxg,Abelev:2013cva}; indeed, the latters are constituted of only newly produced partons and follow the ``antiflow'' behaviour of the directed flow at midrapidity (negative slope), while $K^+$ and $p$ get contributions from the flow of the initial nuclei due to the up and down quarks from which they hadronize.
This source of $v_1$ gives similar contribution to $\pi^+$ and $\pi^-$ so that the splitting due to it is absent or negligible for pions.
This effect is clearly seen in transport simulations with the PHSD model for the $v_1$ of pions and kaons in p+Au collisions at top RHIC energy \cite{Oliva:2019kin} shown and discussed in Sec.~\ref{sec:v1_asym}.
This source of $\Delta v_1$ should be added to that induced by the EMF and the opposite sign between theoretical calculation that do not include this effect \cite{Gursoy:2018yai} and the experimental data \cite{Acharya:2019ijj} may indicate that in symmetric collisions the electromagnetically-induced $v_1$ splitting is subdominant with respect to the contribution from the flow of the initial nuclei (this is not the case for asymmetric collisions as we will discuss in the next section). It can be surprising that this holds even at LHC energies, since the baryon stopping effects are observed to decrease with increasing collision energy, as indicated by the STAR data at different $\sqrt{s_{NN}}$ \cite{Adamczyk:2014ipa,Adamczyk:2017nxg} and by the measurement of a smaller magnitude of $v_1$ at LHC energy \cite{Abelev:2013cva} with respect to lower energies.
Nevertheless, the contribution to the flow from the initial nucleons can be significant, especially in the proton and antiproton $v_1$ splitting as well as in those between charged kaons, therefore influencing the charge dependence of the inclusive hadron $v_1$ in opposite way to the prediction for the EMF effect \cite{Gursoy:2018yai}.
Another possible reason of the discrepancy between theory and experiment for LHC collisions is that the description of the EMF is not enough realistic to account for the tiny unbalance between the magnetic contribution in the Lorentz force and the electric one and one needs to perform a full calculation which accounts for the back-reaction of the accelerated particles on the fields themselves and the pre-equilibrium effects of the field evolution in the early stage.

The ALICE Collaboration presented also the data for the $v_1$ of $D^0$ and $\overline{D}^0$ in the 10--40\% centrality class \cite{Acharya:2019ijj}.
The directed flow of neutral $D$ mesons is shown in Fig.~\ref{fig:v1_eta_LHC}(b): the data favour a positive slope for the pseudorapidity dependence of the $v_1$ of $D^0$ (green circles) and a negative slope for that of $\overline{D}^0$ (orange diamonds) with a significance of about $2\sigma$.
This is different from the observations at top RHIC energy, where a negative slope is found for both particles \cite{Adam:2019wnk}.
Furthermore, the $v_1$ for $D$ mesons is about three orders of magnitude larger than that measured for charged particles shown in panel (a) and this should not be explainable with the different transverse momentum integration ranges.
\\
The theoretical predictions, based on the Langevin approach embedded in a kinetic \cite{Das:2016cwd} or hydrodynamic \cite{Chatterjee:2018lsx} framework, underestimate for about one order of magnitude the ALICE data for the directed flow splitting between $D^0$ and its antiparticle depicted in Fig.~\ref{fig:v1_eta_LHC}(d); the corresponding linear fit extracted by ALICE is
$$
d\Delta v_1/d\eta=\left[4.9\pm1.7\,\rm{(stat.)}\pm0.6\,\rm{(syst.)}\right] \times 10^{-1},
$$
suggesting a positive slope with a significance of $2.7\sigma$.
The theoretical calculations \cite{Das:2016cwd,Chatterjee:2018lsx} predicted a negative slope also at LHC energies, as for the RHIC case.
Hence, the sign of the slope of the $d\Delta v_1^D$ observed by ALICE is in disagreement with both the theoretical expectations and the STAR measurements at $\sqrt{s_{NN}}=200$ GeV \cite{Adam:2019wnk} shown in Fig.~\ref{fig:v1_y_RHIC}(b).
As discussed for the case of charged hadrons, also for the $v_1$ splitting of $D$ mesons the tension between theory and experiment at LHC energy may be related to an underestimated influence of the magnetic field.
Nevertheless, the much larger effect observed for the heavy mesons supports the role attributed to them to be a sensitive probe of the initial EMF produced in ultrarelativistic HICs.

\begin{figure*}[!t]
\centering
\includegraphics[trim={0 0 0 0},clip,width=1.8\columnwidth]{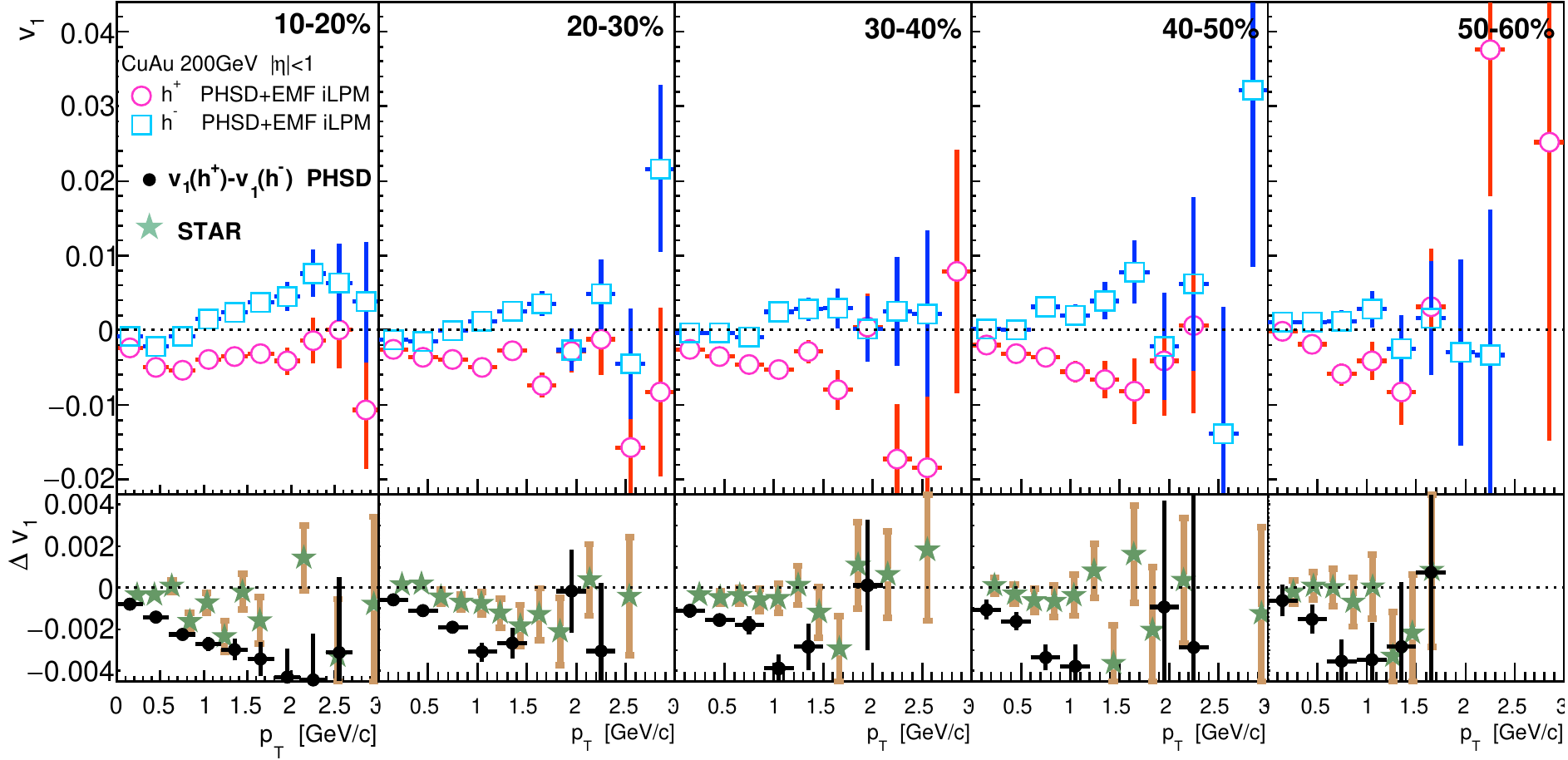}
\caption{Transverse momentum dependence of the directed flow of positively and negatively charged hadrons (upper panels) and their difference $\Delta v_1^{h}=v_1(h^+)-v_1(h^-)$ (lower panels) for Cu+Au collisions at $\sqrt{\sigma_{NN}}=200$ GeV and various centralities computed by the PHSD model in comparison to the experimental data (star markers) from STAR Collaboration \cite{Adamczyk:2016eux,Niida:2016khm}. The figure is taken from Ref.~\cite{Toneev:2016bri}.}
\label{fig:v1_pt_CuAu}
\end{figure*}

\section{Directed flow in asymmetric systems}
\label{sec:v1_asym}

In the previous section we have discussed the theoretical and experimental results for the effect on the directed flow of the EMF produced in symmetric Au+Au and Pb+Pb collisions, where the dominant components are the magnetic field $B_y$ mainly produced by the spectator charges and the electric field $E_x$ generated by Faraday induction due the time decrease of the magnetic field.
In asymmetric systems, e.g. Cu+Au, the larger components are still $B_y$ and $E_x$, but the latter includes a big contribution from the Coulomb field due to the different number of protons inside the colliding nuclei, so that a substantial electric field directed from the heavier to the lighter nucleus is created in the interaction area, as shown in Fig.~\ref{fig:EMF_space}(d).
\\
The EMF produced in Cu+Au collisions along with their influence on the directed flow of light hadrons as a function of rapidity and transverse momentum $p_T$ has been studied with the PHSD approach in Refs.~\cite{Voronyuk:2014rna,Toneev:2016bri}. For collisions at top RHIC energy they found that without the EMF the $v_1$ of charged pions as a function of transverse momentum $p_T$ varies between $0.5-1\%$ in absolute value. The inclusion of the EMF in the simulation splits the distributions of positively and negatively charged pions, pushing the $v_1(\pi^+)$ upward and  $v_1(\pi^-)$ downward with respect to the case without EMF; moreover, the charge splitting $\Delta v_1^{\pi}=v_1(\pi^+)-v_1(\pi^-)$ increases with increasing $p_T$.
\\
In Fig.~\ref{fig:v1_pt_CuAu} we show the PHSD results from Ref.~\cite{Toneev:2016bri} for the transverse momentum dependence of the directed flow of positive ($h^+$) and negative ($h^-$) charged hadrons integrated in the pseudorapidity interval $\vert\eta\vert<1$ for asymmetric Cu+Au collisions at $\sqrt{\sigma_{NN}}=200$ GeV in five centrality intervals.
In the upper panels the $v_1$ of $h^+$ and $h^-$ is plotted separately with magenta circles and blue squares respectively, whereas the lower panels show the difference $\Delta v_1^h=v_1(h^+)-v_1(h^-)$ in comparison to the experimental data from the STAR Collaboration \cite{Adamczyk:2016eux}.
The directed flow splitting of charged hadrons is about 10\% of the $v_1$ magnitude \cite{Adamczyk:2016eux}.
For $p_T<2$ GeV $\Delta v_1^h$ is negative, namely $h^-$ particles have a larger $v_1$ than $h^+$ with respect to the event-plane determined from spectators in the Au-going direction\footnote{In Refs.~\cite{Toneev:2016bri} and \cite{Adamczyk:2016eux} the directions of the initial Au and Cu nuclei are inverted and different conventions for the event-plane determination are used.}.
This is in line with the expectation from the direction of the $x$ component of the electric field which points from the Au nucleus towards the Cu nucleus \cite{Voronyuk:2014rna,Deng:2014uja}; therefore, in the considered reference frame (Cu at $x>0$ moving towards negative rapidity), positive electric charges are pushed upwards and negative charges downwards along the impact parameter direction, generating the observed splitting.
For higher transverse momenta the experimental uncertainties are larger and $\Delta v_{1}$ becomes consistent with zero. It tends to disappear also going towards more peripheral collisions, as visible for the 50\%--60\% centrality class.

\begin{figure*}[t!]
\centering
{\includegraphics[width=0.9\columnwidth]{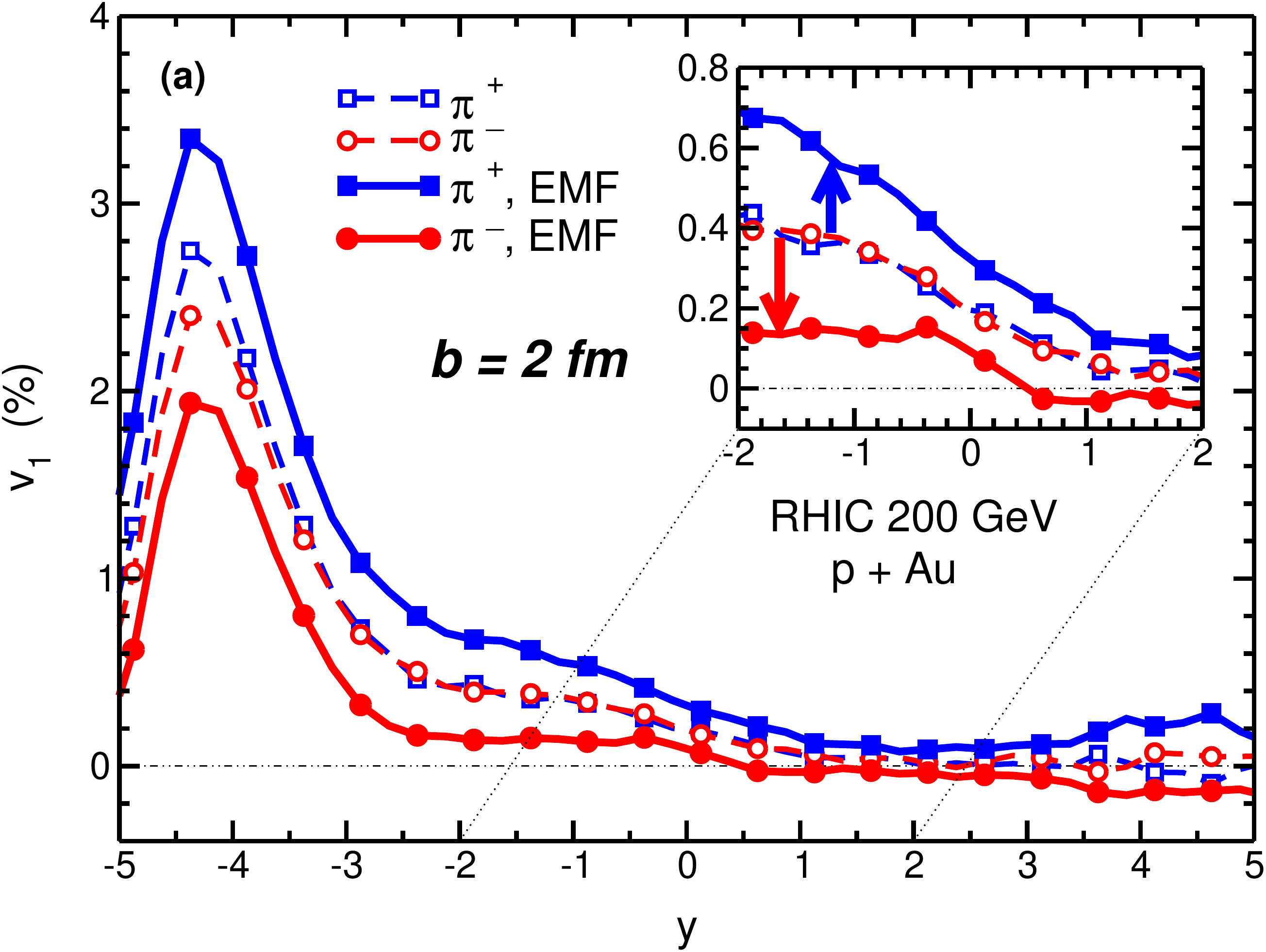}\qquad
\includegraphics[width=0.9\columnwidth]{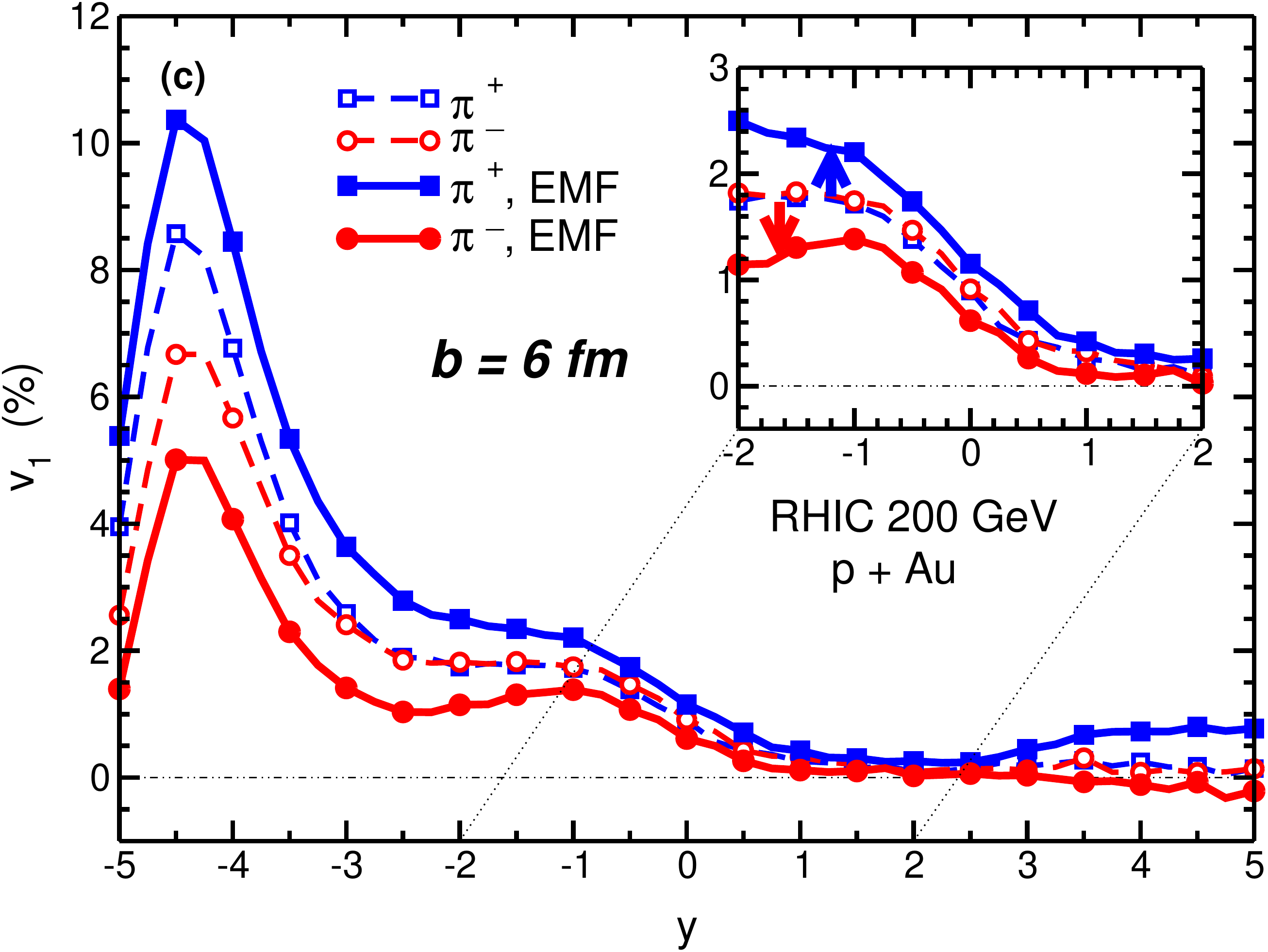}}
\\[\baselineskip]
{\includegraphics[width=0.9\columnwidth]{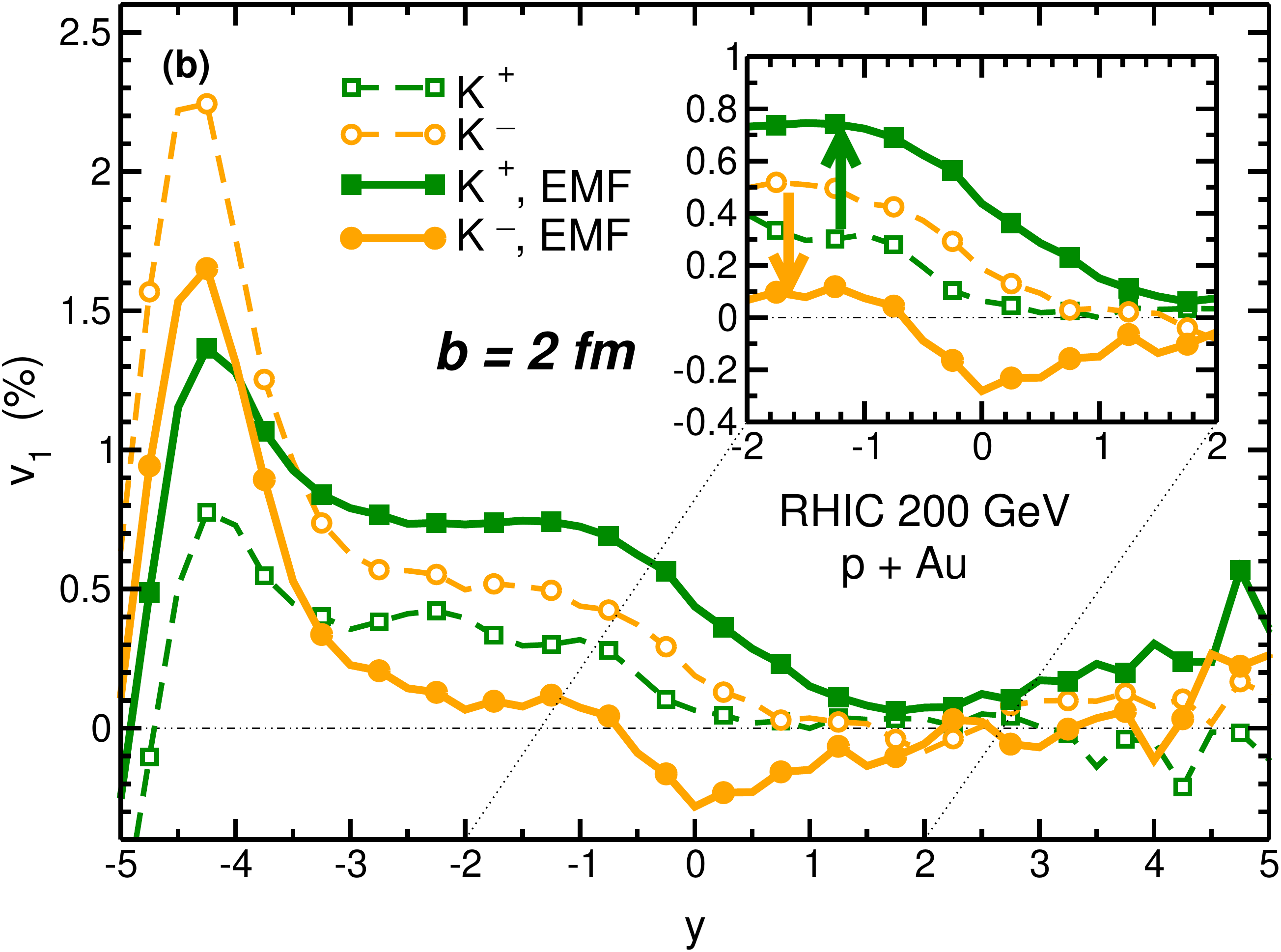}\qquad
\includegraphics[width=0.9\columnwidth]{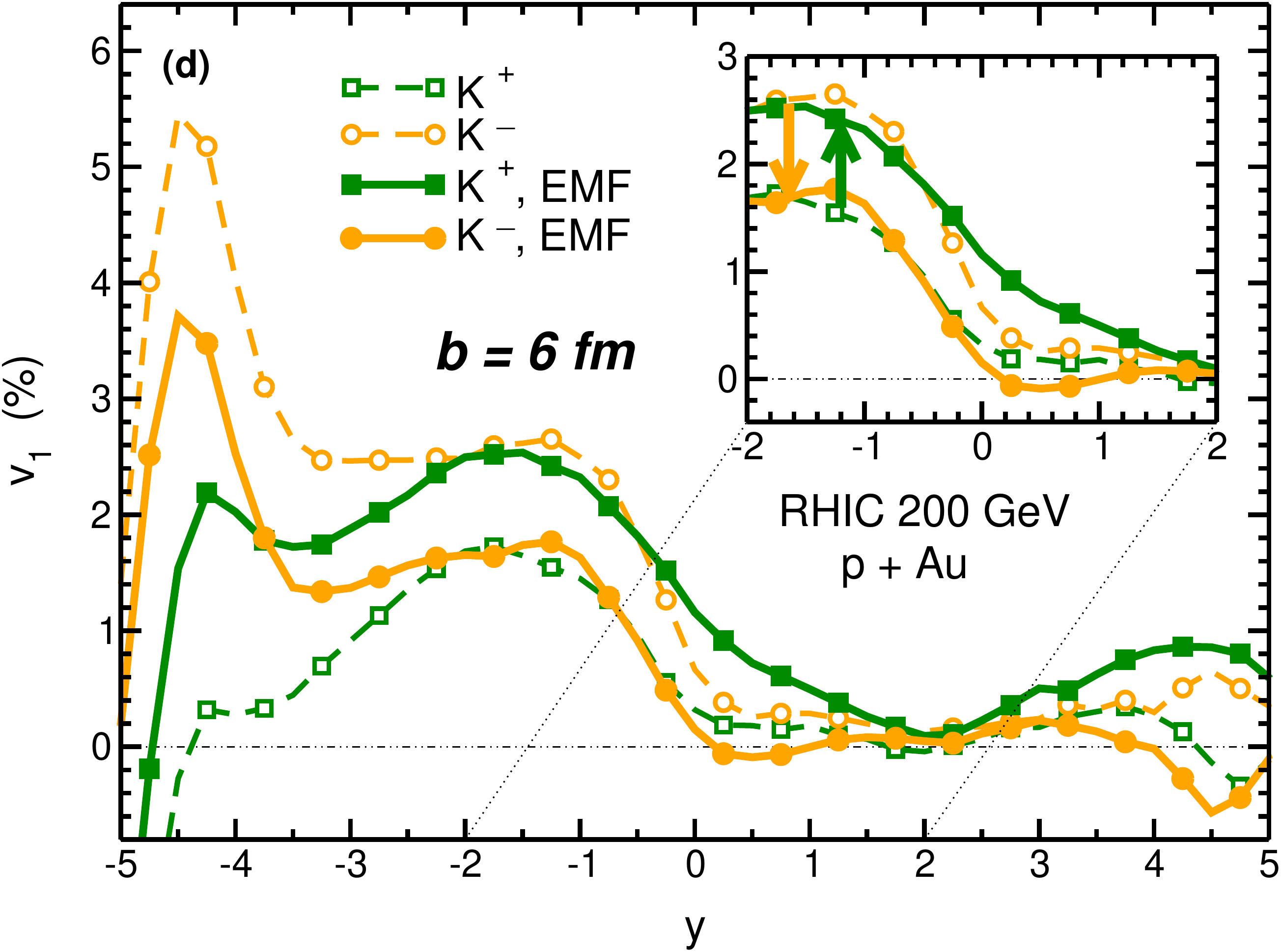}}
\caption{(Color online) Directed flow in percentage of pions (a-c) and kaons (b-d) as a function of rapidity for $b=2$ fm (a-b) and $b=6$ fm (c-d) p+Au collisions at $\sqrt{s_{NN}}=200$ GeV obtained with PHSD simulations with (solid curves) and without (dashed curves) electromagnetic fields. The inset panels are zooms of the rapidity window $\left|y\right|<2$, with arrows highlighting in which direction the presence of the electromagnetic fields affects the $v_1$ observable. The figure is taken from Ref.~\cite{Oliva:2019kin}.}
\label{fig:v1_y_pAu200GeV}
\end{figure*}

Since in asymmetric collisions the Coulomb electric field due to the difference in proton number of the colliding nuclei governs the influence of the EMF effect on particle $v_1$, one can expect that the impact increases going to extremely asymmetric systems, such as proton-nucleus collisions.
The directed flow of light mesons in p+Au collisions at top RHIC energy has been studied, for the first time within the transport framework, by means of the PHSD approach \cite{Oliva:2019kin}.
The effect of the EMF has been disentangled performing simulation with and without the inclusion of those fields and the impact on final observables has been investigated both in minimum bias collisions and for fixed impact parameter.
\\
In Fig.~\ref{fig:v1_y_pAu200GeV} we show the results of Ref.~\cite{Oliva:2019kin} obtained with the PHSD calculations for the rapidity dependence of the $v_1$ of pions (a-c) and kaons (b-d) in p+Au collisions at $\sqrt{s_{NN}}=200$ GeV with impact parameter $b=2$ fm (a-b) and $b=6$ fm (c-d). Along with the complete simulations corresponding to the solid curves, we plot by dashed lines the results without the inclusion of EMF. For each plot a zoom of the $v_1$ at central rapidity is depicted in the inset panel.
In this calculations the incoming proton moves at $x>0$ with positive rapidity and the Au nucleus moving towards negative rapidity has the centre located at $x<0$. Due the big size difference of the two ions, most of the fireball dynamics and particle production takes place at backward rapidity.
\\
Focusing on the directed flow of pions in panels (a) and (c), we see that without EMF $\pi^+$ (dashed blue line with squares) and $\pi^-$ (dashed red line with circles) have the same $v_1$, with mild difference only at rapidity $y\lesssim-3$ as the two particles have different absorption rate in the target region.
The EMF clearly influence the directed flow of pions: $\pi^+$ (solid blue line with squares) are pushed towards the positive $x-$direction and oppositely $\pi^-$ (solid red line with circles) are kicked along the negative $x$; the whole effect is to split the $v_1$ of the pions with opposite electric charge. 
Comparing simulation at $b=2$ fm (a-b) with those at $b=6$ fm (c-d), we see that the push of the EMF is stronger for higher impact parameters; this is comprehensible by looking to the plots of the electric and magnetic fields in Fig.~\ref{fig:EMF_space}(e-f).
\\
We now discuss the directed flow of kaons shown in panels (b) and (d), which is not only influenced by the EMF but keep trace also of the strong vorticity embedded in the system.
From the dashed curves, we see that $K^+$ (green line with squares) and $K^-$ (orange line with circles) show a splitting in the $v_1$ even in simulations in which the EMF are switched off.
This interesting effect, already discussed in Sec.~\ref{sec:v1_sym}, can be linked to the intense angular momentum owned by the two impinging nuclei, that is strongly kept by the $u$ and $d$ quarks inside the initial nucleons and partly transferred to the produced hot QGP.
The huge angular momentum of the initial partons is visible in $K^+$ ($\overline{s}u$) which receive more contributions with respect to $K^-$ ($s\overline{u}$) from the sidewards deflection of the heavy nucleus \cite{Dunlop:2011cf,Adamczyk:2017nxg}; as a consequence, $K^+$ has a smaller $v_1$ with respect to $K^-$ at backward rapidity, that is the Au-going direction.
However, accounting for the EMF (solid lines) the distribution of $K^+$ is pushed upward and that of $K^-$ is pushed downward at $y<0$; this electromagnetically-induced splitting dominates over that generated by the initial vorticity of the system, so that the final result presents the same trend as seen for the pions.
\\
The role of the EMF on the charge-dependent $v_1$ in proton-nucleus collisions can be understood in a similar way as for the Cu+Au colliding system.
For both pions and kaons, the sign of the splitting in $v_1$ results from the balance between sideways pushes on charged particles by the electric and magnetic part of the Lorentz force \eqref{eq:lorentz} \cite{Voronyuk:2014rna,Toneev:2016bri,Gursoy:2014aka,Gursoy:2018yai,Das:2016cwd,Coci:2019nyr,Chatterjee:2018lsx}. We can distinguish two contributions in the electric field: the Faraday induction due to the decrease of the magnetic field and the Coulomb interaction due to the difference in proton number of the colliding system.
The force exerted by the Faraday-induced $E_x$ has opposite directions at forward and backward rapidity and goes in both regions against the push  of the magnetic field $B_y$. The $x$-projected Coulomb electric field, instead, has the same direction at both positive and negative $y$.
In the backward hemisphere, where the main QGP production occurs, Coulomb and Faraday electric fields sum up: according to our convention for the reference frame, the electric part of the Lorentz force pushes positively (negatively) charged particles along the positive (negative) $x$ direction while the magnetic contribution has the opposite impact.
As highlighted by the arrows drawn in the insets of Fig.~\ref{fig:v1_y_pAu200GeV}, the winner of this force competition in p+Au reactions is $E_x$, attracting $\pi^-$ and $K^-$ towards the beam axis and pushing far $\pi^+$ and $K^+$.
The splitting increases for higher (absolute) values of rapidity and for larger impact parameters.
Furthermore, it depends also on particle species, having a stronger impact on kaons with respect to pions. 
This can be related to the different mass and consequently different velocity of the two
meson species: slower particles (the kaons) feel longer the influence of the EMF and undergo to a smaller force from the magnetic field ($F_{B}\propto v\times B$) that leads to a smaller compensation of the electric field push.

\section{Discussion and conclusions}

In this review the main approaches adopted in theoretical calculations for describing the generation and time evolution of the electromagnetic fields (EMF) in ultrarelativistic nuclear collisions has been discussed, highlighting the significant differences in the fields produced in symmetric and asymmetric systems. Indeed, in the latter case, the different number of protons of the two colliding nuclei generates a more inhomogeneous distribution of the EMF and give rise to substantial values of some components relatively smaller in the symmetric case. This asymmetry is taken to its extreme in the case of proton-nucleus collisions, where the profiles of the fields are almost completely determined by the heavy ion.
\\
The EMF drive many interesting effects, some of which are also affected by the fascinating early-time dynamics of nuclear collisions when the EMF attain their maximal strength and the system is in a very anisotropic and out-of-equilibrium state characterized also by the intense vorticity induced by the huge angular momentum owned by the colliding nuclei.
\\
The presence of the EMF breaks the spherical symmetry in the plane perpendicular to the beam direction and induces an azimuthal asymmetry in the final particle distribution. In particular, due to the combined geometry of EMF and QGP expansion, the directed flow $v_1$ has been considered as the most promising flow harmonics to see the impact of the EMF in heavy-ion collisions and small colliding systems. Its magnitude is determined also by the initial-state fluctuations and the QGP vortical patterns, but its charge-dependent behaviour gives direct access to the EMF strength and decay rate, which in turn give also information on the presence of electric charges in the early-stage as well as the electromagnetic response of the QGP.

The recent theoretical and experimental results on the directed flow of light hadrons and heavy mesons has been reviewed.
Kinetic calculations of the directed flow of light mesons in p+Au collisions at top RHIC energy clearly show the importance for the numerical codes to be able to capture the effect of the EMF on the $v_1$ but also to do not miss other sources of $v_1$ such as the vorticity of the medium and the baryon stopping mechanism, which act in different way on some particles (e.g., $K^+$ and $p$) with respect to their antiparticles ($K^-$ and $\overline{p}$), hence inducing a charge-dependent $v_1$ which has to be combined to the electromagnetically-induced splitting.
In heavy-ion collisions a much larger effect on the neutral $D$ mesons with respect to light particles has been predicted by the theoretical models for the $v_1$ generated by the EMF and the initial tilt of the fireball; the recent experimental observations support the role attributed to the heavy flavour to be a sensitive probe of the early dynamics of ultrarelativistic collisions, due to their strong sensitivity to the vorticity and the EMF there produced.
However, the recent experimental measurements of the $v_1$ of $D^0$ and $\overline{D}^0$ at LHC energies cannot be explained by the current theoretical calculations and show opposite behaviour with respect to the data in RHIC collisions, suggesting that at the highest energies there may be a not trivial and not yet understood interplay between the initial-state effects and the EMF; this could lead to a dominant impact of the magnetic field over those caused by the electric field and by the fireball vorticity.
It would be interesting to see if the trend observed at LHC in symmetric Pb+Pb collisions is maintained in asymmetric systems (e.g., p+Pb) where the huge electric field should oppose more to the magnetic field.

More efforts are needed both on the theoretical and the experimental side to disentangle the role of the EMF from the other sources on the generation of a charge-odd directed flow.
A more detailed scan of various systems and energies is desirable, from heavy-ion collisions to proton-induced reactions at the different beam energies.
The current tension between theory and experiment makes clear the importance to perform precise measurements and to further develop the numerical codes able to describe all the phenomena taking place especially in the early-stage of the collisions.

\section*{Acknowledgements}
The author appreciates very useful discussions with Elena Bratkovskaya, Wolfgang Cassing, Salvatore Plumari, Marco Ruggieri, Olga Soloveva, Vadim Voronyuk.
The author is financially funded by the Alexander von Humboldt-Stiftung and acknowledges support from the COST Action THOR CA15213 and from the Deutsche Forschungsgemeinschaft (DFG) through the grant CRC-TR 211 'Strong-interaction matter under extreme conditions'.

\bibliography{References}
\bibliographystyle{epj}

\end{document}